\documentclass[aps,groupedaddress,twocolumn,floatfix,showpacs,tightenlines,superscriptaddress]{revtex4-1}
\usepackage{graphicx}
\usepackage{float}
\usepackage{color}
\usepackage{psfrag}
\usepackage{dcolumn}
\usepackage{bm}

\usepackage{amssymb}
\usepackage{amsmath}
\usepackage{amsfonts}
\usepackage{longtable}
\usepackage{xspace}
\usepackage{mathrsfs}
\usepackage{gensymb}
\usepackage[table]{xcolor}

\newcommand{\beq}{\begin{equation}}
\newcommand{\eeq}{\end{equation}}
\newcommand{\bea}{\begin{eqnarray}}
\newcommand{\eea}{\end{eqnarray}}
\newcommand{\bes}{\begin{subequations}}
\newcommand{\ees}{\end{subequations}}

\newcommand {\sth}{\sin\theta}
\newcommand {\ssth}{\sin^2\theta}
\newcommand {\cth}{\cos\theta}
\newcommand {\ccth}{\cos^2\theta}

\allowdisplaybreaks

\begin{document}
\title{Local and Approximate Classification of Spacetimes in the Transverse Frames}

\author{Nicole Rosato}
\affiliation{Center for Computational Relativity and Gravitation,\\
School of Mathematical Sciences, Rochester Institute of Technology, 85 Lomb Memorial Drive, Rochester, New York 14623}

\author{Hiroyuki Nakano} 
\affiliation{Center for Computational Relativity and Gravitation,\\
School of Mathematical Sciences, Rochester Institute of Technology, 85 Lomb Memorial Drive, Rochester, New York 14623}
\affiliation{Faculty of Law, Ryukoku University,\\ 
67 Tsukamoto-cho, Fukakusa Fushimi-ku, Kyoto 612-8577, Japan
}

\author{Carlos O. Lousto}
\affiliation{Center for Computational Relativity and Gravitation,\\
School of Mathematical Sciences, Rochester Institute of Technology, 85 Lomb Memorial Drive, Rochester, New York 14623}

\date{\today}

\begin{abstract}
We revisit the definition of transverse frames and tetrad choices with regards to its application to numerically generated spacetimes, in particular those from the merger of binary black holes. We introduce the concept of local and approximate algebraic Petrov types in the strong field regime. We define an index $\mathcal{D}=\sqrt{12/I}\left(\Psi_2 - \Psi_3^2/\Psi_4\right)$ able to discriminate between Petrov types II and D and define regions of spacetime of those approximate types when used in conjunction with the speciality invariant $S=27J^2/I^3$. We provide an explicit example applying this method to Brill-Lindquist initial data corresponding to two nonspinning black holes from rest at a given initial separation. We find a doughnut-like region that is approximately of Petrov type II surrounded by an approximately Petrov type D region.
We complete the study by proposing a totally symmetric tetrad fixing of the transverse frame that can be simply implemented in numerically generated spacetimes through the computation of spin coefficients ratios. We provide an application by explicitly deriving the Kerr-perturbative equations in this tetrad. 
\end{abstract}
\pacs{04.25.dg, 04.25.Nx, 04.30.Db, 04.70.Bw}

\maketitle

\section{Introduction}\label{sec:intro}

With the new generation of gravitational wave detectors on the horizon, such as the space-based detector LISA and the constant advancements to the LIGO and VIRGO detectors, it is of interest to develop more accurate, less computationally expensive methods of extracting gravitational wave signals from binary compact object mergers.
Asymptotically, the spacetime can be rotated into a frame that pushes direct physical information into the Weyl scalar, $\Psi_4$, that we use to calculate outgoing gravitational radiation. This, in turn, allows the computation of gravitational wave strain $h$ 
\begin{equation}
    \Psi_4 = - \ddot{h}^+ + i\ddot{h}^\times =\ddot{h} \,,
\end{equation}
which implies
\begin{equation}
    h = -\int_{-\infty}^{t} \int_{-\infty}^{t'}\Psi_4 \,dt''dt' \,,
\end{equation}
and is directly related to the measures of the gravitational wave detectors.

Teukolsky~\cite{Teukolsky:1973ha}, in 1972, presented separable equations for the radiative parts of electromagnetic and gravitational perturbations on a Kerr background in Boyer-Lindquist coordinates, derived from a formalism introduced by Newman and Penrose in 1962~\cite{Newman62a}.
This Newman-Penrose formalism uses the Weyl tensor $C_{\alpha\beta\gamma\delta}$
which is the trace-free component of the Riemann curvature tensor $R_{\alpha\beta\gamma\delta}$, to develop a formalism that re-expresses the 10 independent components of the Weyl tensor in terms of five complex scalars $\Psi_0,\,\cdots,\, \Psi_4$, the Newman-Penrose (Weyl) scalars. In a matter-free spacetime, asymptotically $\Psi_0,\,\Psi_2$ and $\Psi_4$ represent ingoing radiation, the Coulomb field, and outgoing radiation, respectively. Mathematically, they are formed by contraction of the Weyl tensor with an arbitrary (complex) null tetrad $(l^\alpha,\, n^\alpha,\, m^\alpha,\, \bar m^\alpha)$. The tetrad itself is formed from combinations of also arbitrary, but orthonormal basis vectors. The Weyl tensor and associated tetrad vectors are contracted in the following way:
\begin{align}
\Psi_0 &= C_{\alpha\beta\gamma\delta}\,l^\alpha m^\beta l^\gamma m^\delta \,,\cr
\Psi_1 &= C_{\alpha\beta\gamma\delta}\,l^\alpha n^\beta l^\gamma m^\delta \,,\cr
\Psi_2 &= C_{\alpha\beta\gamma\delta}\,l^\alpha m^\beta \bar m^\gamma n^\delta \,,\cr
\Psi_3 &= C_{\alpha\beta\gamma\delta}\,l^\alpha n^\beta \bar m^\gamma n^\delta \,,\cr
\Psi_4 &= C_{\alpha\beta\gamma\delta}\,n^\alpha \bar m^\beta n^\gamma \bar m^\delta \,.
\label{weylscalars}
\end{align}
The constraints on the tetrad are that it must satisfy the relationships $l^\alpha n_\alpha=-1$ and $m^\alpha \bar m_\alpha=1$, and have all other inner products vanishing. 

Teukolsky uses this formalism to specify a null tetrad, which differs from the background Kinnersley tetrad~\cite{Kinnersley_1969} (in which the only nonvanishing Weyl scalar is $\Psi_2$), by leading order perturbations. The Kinnersley tetrad has $\{l^\alpha\}$ and $\{n^\alpha\}$
vectors 
along the two principal null directions (PNDs)~\cite{Kramer80}. 

At late times in a binary black hole (BBH) merger, Teukolsky's perturbed Kinnersley tetrad is expected to have $\Psi_1=\Psi_3=0$, which is the condition that characterizes a transverse frame. The Teukolsky formalism requires classification of a tetrad into a specific transverse frame. There are three possible transverse frames in any generic spacetime and choosing among them is nontrivial. 

A ``quasi-''Kinnersley (QK) frame is transverse, which means the Weyl scalars $\Psi_1=\Psi_3=0$, and since they are nonphysical in a matter-free spacetime, this means that more physical information is encoded in the gravitational wave scalar $\Psi_4$, the ingoing radiation scalar $\Psi_0$ and the Coulomb field $\Psi_2$.

The QK frame has since been explored in greater detail in a series of articles by Beetle, Bruni, Burko, and Nerozzi (Refs.~\cite{Beetle:2004wu}, \cite{Nerozzi:2004wv}, \cite{Burko:2005fa}, and \cite{Nerozzi:2005hz}). The first paper extends the analytic work done in Ref.~\cite{Campanelli:2005ia} to the numerical regime, by outlining the limits of the QK frame, and establishing methodologies for incorporating it into full numerical relativistic simulations. 
Beetle et al.~\cite{Beetle:2004wu} propose the construction of a set of transverse null tetrads that are equivalent under spin-boost and exchange transformations. However, only one frame will be QK, and explicit rules for its identification are presented in the paper. We will use their definitions extensively in the coming sections.

The first paper defines the framework for Ref.~\cite{Nerozzi:2004wv} which performs analytic rotations of the five Weyl scalars $\Psi_0,\,\cdots,\,\Psi_4$ using Type I and Type II rotations, and solves for two constant, complex parameters $\bar a$ and $b$ necessary to rotate into a QK frame (the overbar signifies the complex conjugate of $a$). The Type I and Type II rotations of the Weyl scalars in the original frame lead to two equations for $\Psi''_1$ and $\Psi''_3$ which can be solved for $\bar a$ and $b$ (see Eqs.~(\ref{eq:TypeI}) and (\ref{eq:TypeII}) below). The double apostrophe delineates a doubly transformed scalar. The polynomial $\Psi''_1$ is sixth order in $\bar a$, but Nerozzi et al. provide a method of reducing it to fourth order.
When the spacetime is perturbatively close to Kerr, namely at late times (post-merger) or at distances far from the binary ($r\to\infty$), the QK frame approximates the background Kinnersley frame. The correct transverse frame is chosen by identifying the eigenvalues of the Weyl tensor (i.e., of Eq.~(\ref{qmat}) below) that approach $2\Psi_2$ in a transverse frame \cite{Nerozzi:2004wv}. 
The authors go on to discuss how to differentiate between the QK frame and other transverse frames, they introduce a new curvature scalar $\xi=\Psi_0\Psi_4$, and finally provide analytic rotations into the QK frame of the Weyl scalars in algebraically special spacetimes. 
Using the Newman-Penrose formalism, the authors provide a fully analytic prescription for constructing the three transverse frames and identifying the one that is QK.

Finally, Ref.~\cite{Burko:2005fa} goes further in-depth on the Beetle-Burko radiation scalar, $\xi$, the curvature invariant for general relativistic spacetimes. The scalar measures the total amount of radiation --- ingoing, outgoing, and spurious, in a spacetime. The paper applies it to a number of initial data sets describing single black-hole spacetimes. 
The analysis is done entirely in the QK frame, identified by demanding continuity in $\Psi_2$ and $\xi\to0$ as $r\to\infty$.


The transverse frame is used in a number of different analytical applications. For instance, it was used in the Lazarus project~\cite{Baker:2001sf,Campanelli:2005ia}. This work constructs Cauchy data for the Teukolsky evolution~\cite{Baker:2001sf} and then Ref.~\cite{Campanelli:2005ia} rotates the resultant scalars into the QK frame, which allows for the extraction of information about the background Kerr solution and, in turn, the gravitational radiation. We are also interested in using the associated non-QK, but still transverse, frames to analyze and classify the spacetime very close to the black holes.

In 2001, the Lazarus project~\cite{Baker:2001sf} showed that late-time evolutions of a BBH spacetime can be seen as a perturbed Kerr spacetime, and used this to extract information about gravitational radiation. In 2006, the second iteration of this project~\cite{Campanelli:2005ia} sought to improve late-time gravitational wave extraction further by rotation of a spacetime into a QK frame, and then explored the late-time behavior of a merged BBH system, which should differ only perturbatively from Kerr. 
A recent paper~\cite{Iozzo:2020jcu} displays the use of the full set of Weyl scalars to extract accurate gravitational waves asymptotically in numerically generated spacetimes.

In this paper, we apply the techniques of Ref.~\cite{Nerozzi:2004wv} to analytic initial data of black hole pairs in order to rotate a particular spacetime into a transverse frame that is QK far from the binary, and generically transverse close to them. We specifically study the strong-field regime of these spacetimes and classify interesting regions into different Petrov types. We also present a new index $\mathcal{D}$ that, when used in conjunction with the Baker-Campanelli~\cite{Baker:2000zm} speciality invariant, 
\begin{equation}\label{eq:S}
S=\dfrac{27J^2}{I^3} \,,
\end{equation}
will allow us to differentiate between Petrov types D and II in the strong field regime where there is no a priori knowledge of the spacetime's classification.

In Section~\ref{sec:TF}, first, we briefly review and discuss the analytic method presented by Ref.~\cite{Nerozzi:2004wv} for rotation into the QK frame. Then, in Section~\ref{sec:Class} we show that the QK frame can be used to classify the spacetime close to the black holes, and then discuss where this definition breaks down. 
As a consequence, we construct the $\mathcal{D}$-index and use it, in conjunction with the $S$-invariant and a transverse frame (other than the QK frame), to successfully classify the strong-field region of a set of BBHs that uses Brill-Lindquist initial data. 
In Section~\ref{sec:tetrad}, we proceed to completely fix the tetrad from a transverse frame in a fashion that is simple to implement in fully numerically generated spacetimes, products of the dynamical evolution of BBHs and its final merger.
Finally, in Section~\ref{sec:disc}, we summarize our analysis
and discuss some applications.


\section{Transverse Frames determination}\label{sec:TF}

We start by reviewing some of the key elements needed to our discussion of the (local) classification of spacetimes and to fix notation and will correct and discuss some typos/issues found in the literature.

\subsection{Analytic Null Rotation into the Quasi-Kinnersley Frame}
\label{sec:qkintro}

To rotate into the QK frame, begin with any arbitrary frame $\mathcal{F}$ characterized by a set of arbitrary null tetrad vectors 
$(l^\alpha,\, n^\alpha,\, m^\alpha,\, \bar m^\alpha)$. From the tetrad vectors, the Weyl scalars in $\mathcal{F}$ can be built up via Eq.~(\ref{weylscalars}).
Then, one can write the eigenvalue ($\lambda$) equation associated with the Weyl tensor (as in Ref.~\cite{Kramer80})
\begin{equation}\label{eig}
\frac{1}{2} C_{\alpha\beta\mu\nu} X^{\mu\nu} = \lambda X^{\alpha\beta} \,,
\end{equation}
where $C_{\alpha\beta\mu\nu}$ is the Weyl tensor and $X^{\alpha\beta}$ is an associated eigenbivector. 
This can be reduced to 
\begin{equation}
Q_{ab}r^b = \lambda r_a \,.
\end{equation}
The complex, symmetric matrix $Q_{ab}$ takes the form
\begin{equation}\label{qmat}\centering
Q_{ab}=
\begin{bmatrix}
\Psi_2 - \displaystyle{\frac{\Psi_0 + \Psi_4}{2}} & \displaystyle{\frac{i(\Psi_4 - \Psi_0)}{2}} & \Psi_1 - \Psi_3 \\
\displaystyle{\frac{i(\Psi_4 - \Psi_0)}{2}} &\Psi_2 + \displaystyle{\frac{\Psi_0 + \Psi_4}{2}} & i(\Psi_1 + \Psi_3) \\
\Psi_1 - \Psi_3 & i(\Psi_1 + \Psi_3) & - 2\Psi_2 
\end{bmatrix} \,.
\end{equation}
In vector notation, this is
\begin{equation}
Q\textbf{r} = \lambda\textbf{r} \,.
\end{equation}
The eigenvalues $\lambda$ of $Q$ are found by setting
\begin{equation}\label{det}
\text{det}|Q-\lambda I_3| = 0 \,,
\end{equation}
and solving for $\lambda$. Here, $I_3$ denotes the $3\times3$ identity matrix.
For a general 4D spacetime, $Q \in \mathbb{C}^{3\times3}$, which means that there are 3 complex eigenvalues $\lambda_1,\,\lambda_2$, and $\lambda_3$. The characteristic polynomial to solve is 
\begin{equation}\label{charpoly}
\lambda^3 - I \lambda + 2 J = 0 \,,
\end{equation}
where $I$ and $J$ are spacetime invariants which can be written in terms of the Weyl scalars as 
\begin{equation}\label{Iinv}
I = \Psi_4 \Psi_0 - 4 \Psi_3 \Psi_1 + 3 \Psi_2^2 \,,
\end{equation}
and
\begin{equation}\label{Jinv}
J=\text{det}
\begin{vmatrix}
\Psi_4 & \Psi_3 &  \Psi_2\\
\Psi_3 & \Psi_2 & \Psi_1 \\
\Psi_2 & \Psi_1 & \Psi_0
\end{vmatrix} \,.
\end{equation} 
All three eigenvalues of $Q_{ab}$ are associated with their own individual transverse frame, in which the Weyl scalars adhere to $\Psi_1 = \Psi_3=0$. There will be one eigenvalue ($\lambda_{P}=\lambda_{\rm QK}$) considered to be principal and specifically associated with the QK frame. Due to the complex nature of the roots of Eq.~(\ref{charpoly}), choosing the principal eigenvalue is nuanced and nontrivial --- continuity must be forced in the strong field region~\cite{Beetle:2004wu}.
A discussion on methods of choosing $\lambda_P$ will proceed in Section~\ref{lambda1}. Once the principal eigenvalue is chosen at every point in the 3D space, it is used to construct the rotation parameter $\bar a$. This is used to transform the Weyl scalars by a Type I rotation:
\begin{align}
&\Psi_0' \to \Psi_0 \,,\cr 
&\Psi_1' \to \Psi_1 + \bar a \Psi_0 \,,\cr 
&\Psi_2' \to \Psi_2 + 2\bar a \Psi_1 + \bar a^2\Psi_0 \,,\cr 
&\Psi_3' \to \Psi_3 + 3\bar a \Psi_2 + 3\bar a^2\Psi_1 + \bar a^3\Psi_0 \,,\cr 
&\Psi_4' \to \Psi_4 + 4\bar a \Psi_3 + 6\bar a^2\Psi_2 + 4\bar a^3\Psi_1 + \bar a^4\Psi_0 \,. 
\label{eq:TypeI}
\end{align} 
Then, rotation parameter $b$ is constructed from $\bar a$, and a Type II rotation is performed via:
\begin{align}
&\Psi_0' \to \Psi_0 + 4b \Psi_1 + 6b^2\Psi_2 + 4b^3\Psi_3 + b^4\Psi_4 \,,\cr 
&\Psi_1' \to \Psi_1 + 3b \Psi_2 + 3b^2\Psi_3 + b^3\Psi_4 \,,\cr 
&\Psi_2' \to \Psi_2 + 2b \Psi_3 + b^2\Psi_4 \,,\cr 
&\Psi_3' \to \Psi_3 + b \Psi_4 \,,\cr 
&\Psi_4' \to \Psi_4 \,. 
\label{eq:TypeII}
\end{align} 

\subsection{Finding the Principal Eigenvalue}\label{lambda1}

Each eigenvalue of $Q_{ab}$ in Eq.~(\ref{qmat}) corresponds to a reference frame from which the Weyl scalars $\Psi_a$ can be computed. One can freely rotate among frames using constants $\bar a$ and $b$, and all such frames constructed using the eigenvalues $\lambda$ are transverse. One specific frame, the QK frame, is associated with an eigenvalue $\lambda_P$ of $Q_{ab}$ that we will consider to be principal. 
The identification of the principal eigenvalue is trivial in this asymptotic region --- it needs only to be twice the magnitude of each of the other two eigenvalues (see Ref.~\cite{Nerozzi:2004wv}). One would like this principal eigenvalue to be at least $C^1$ over the whole spacetime, but the invariants used to construct the eigenvalues are complex, and the eigenvalues themselves contain complex cube roots. These roots introduce branch cuts in the eigenvalues unless the principal is \textit{forced} to be continuous at each point in space in either real or imaginary part. This condition requires that $\lambda_P$ move out of the QK frame and into an alternate transverse frame as $r\to0$.

Notice that since Eq.~(\ref{charpoly}) is only a cubic equation, there exists a fully analytic solution for the three eigenvalues of $Q_{ab}$~\cite{Nerozzi:2004wv}
\begin{align}\label{eq:lambdas}
\lambda_1 &= - \left(P+\frac{I}{3P} \right) \,,\\
\lambda_2 &= - \left(e^\frac{4 \pi i}{3} P+ e^\frac{2 \pi i}{3}\frac{I}{3P}\right) \,,\\
\lambda_3 &= - \left(e^\frac{2 \pi i}{3} P+ e^\frac{4 \pi i}{3}\frac{I}{3P}\right) \,,
\end{align}
where 
\begin{equation}
P = \left[ J + \sqrt{J^2 - \left(\frac{I}{3}\right)^3}\right]^{1/3} \,.
\end{equation}
We are now left with the task of choosing which eigenvalue, $\lambda_{P}$, is QK at every point. This will correspond with $\lambda_{1}$ asymptotically, but it is not necessarily true that $\lambda_{P}=\lambda_{1}$ in the strong-field region.
We have already discussed how, asymptotically, $\lambda_{P}$ should be twice the magnitude of either of the other eigenvalues~\cite{Nerozzi:2004wv,Beetle:2004wu}. Explicitly, it can be said that as $r\to\infty$, $\lambda_{P} = \text{max}_k|\lambda_k|$. However, in the strong-field region, we will show that this definition breaks down.

\subsection{Calculation of Rotation Parameters $\bar a$ and $b$}\label{QKrotation}

Since, in Section~\ref{lambda1}, we outlined an analytic methodology of choosing $\lambda_{P}$ at each point in space for a particular time-slice, we are now ready to rotate the Weyl scalars into a QK frame.  This is done by performing a Type I rotation and then a Type II rotation on the scalars using rotation parameters $\bar a$ and $b$. 
To do this, Ref.~\cite{Nerozzi:2004wv} suggests setting up two equations for the two unknowns by rotating both $\Psi_1$ and $\Psi_3$ into a transverse frame as follows
\begin{align}\label{eqs3132}
\Psi_3& + 3\bar a \Psi_2 + 3\bar a^2\Psi_1 + \bar a^3\Psi_0 \cr
& + b(\Psi_4 + 4\bar a \Psi_3 + 6\bar a^2\Psi_2 + 4\bar a^3\Psi_1 + \bar a^4\Psi_0 ) = 0 \,,\\
\Psi_1 &+ \bar a \Psi_0 + 3b(\Psi_2 + 2\bar a \Psi_1 + \bar a^2\Psi_0) \cr
& + 3b^2(\Psi_3 + 3\bar a \Psi_2 + 3\bar a^2\Psi_1 + \bar a^3\Psi_0) \cr 
& + b^3 (\Psi_4 + 4\bar a \Psi_3 + 6\bar a^2\Psi_2 + 4\bar a^3\Psi_1 + \bar a^4\Psi_0 ) = 0 \,.
\end{align}
Then $b$ can be written as a function of $\bar a$ as follows
\begin{equation}
b = -\frac{\Psi_3 + 3\bar a \Psi_2 + 3\bar a^2\Psi_1 + \bar a^3\Psi_0}{\Psi_4 + 4\bar a \Psi_3 + 6\bar a^2\Psi_2 + 4\bar a^3\Psi_1 + \bar a^4\Psi_0} \,,
\end{equation}
and all that is necessary to do is to find $\bar a$. Equation~(\ref{eqs3132}) provides a sixth order polynomial to be solved for $\bar a$
\begin{align}\label{sixor}
& A_{(1)} \bar a^6 + A_{(2)} \bar a^5 +A_{(3)}\bar a^4+A_{(4)}\bar a^3+A_{(5)}\bar a^2
\cr & +A_{(6)}\bar a+A_{(7)} = 0 \,,
\end{align}
with coefficients
\begin{align}
A_{(1)} &=-\Psi_3\Psi_0^2 - 2\Psi_1^3 + 3\Psi_2\Psi_1\Psi_0 \,,\cr
A_{(2)} &=-2\Psi_3\Psi_1\Psi_0 - \Psi_0^2\Psi_4 + 9\Psi_2^2\Psi_0 - 6\Psi_2\Psi_1^2 \,,\cr
A_{(3)} &=-5\Psi_1\Psi_4\Psi_0 - 10\Psi_3\Psi_1^2 + 15\Psi_3\Psi_2\Psi_0 \,,\cr
A_{(4)} &=-10\Psi_4\Psi_1^2 + 10\Psi_3^2\Psi_0 \,,\cr
A_{(5)} &=5\Psi_3\Psi_0\Psi_4 + 10 \Psi_1\Psi_3^2 - 15\Psi_1\Psi_2\Psi_4 \,,\cr
A_{(6)} &=2 \Psi_3\Psi_1\Psi_4 + \Psi_4^2\Psi_0 - 9\Psi_2^2\Psi_4 + 6 \Psi_2\Psi_3^2 \,,\cr
A_{(7)} &=\Psi_1\Psi_4^2 + 2 \Psi_3^3 - 3 \Psi_2\Psi_3\Psi_4 \,,
\end{align}
which can only be solved using numerical methods. Therefore, it is necessary to find which root, of the six, is associated with the QK frame. Due to the computational complexity of solving a sixth-order polynomial, Nerozzi et al. reduce the polynomial order of Eq.~(\ref{sixor}) to fourth order.

\subsubsection{Reduction to fourth order}\label{reduceto4}

The authors of Ref.~\cite{Nerozzi:2004wv} begin by rotating an arbitrary tetrad so that $n$ (or $l$) is a principal null direction and $\Psi_4$ ($\Psi_0$) vanishes. We will outline their method here, as its result will be useful to us later on. Begin by performing a Type I rotation on $\Psi_4$, and setting it to zero: 
\begin{equation}
b^4 \Psi_4 + 4b^3 \Psi_3 + 6 b^2 \Psi_2 + 4 b \Psi_1 + \Psi_0=0 \,. 
\end{equation}
This can be reduced to a depressed quartic by making the substitution $\hat z = \Psi_4 b + \Psi_3$:
\begin{equation}\label{dquar}
\hat z^4 + 6 \hat H \hat z^2 + 4 \hat G \hat z + \hat K = 0 \,,
\end{equation}
with 
\begin{align}
\hat H &= \Psi_4 \Psi_2 - \Psi_3^2 \,,\cr
\hat G &=\Psi_4^2\Psi_1 - 3 \Psi_4\Psi_3\Psi_2 + 2 \Psi_3^3 \,,\cr
\hat K &= \Psi_4^2 I - 3 \hat H^2 \,, 
\end{align}
where the variables
\begin{align}
\hat \alpha^2 &= 2 \Psi_4\lambda_1 - 4 \hat H \,,\cr
\hat \beta^2 &= 2 \Psi_4\lambda_2 - 4 \hat H \,,\cr
\hat \gamma^2 &= 2 \Psi_4\lambda_3 - 4 \hat H \,,
\end{align}
can be combined so that
\begin{equation}
\hat\alpha \hat\beta \hat\gamma = 4 \hat G \,.
\end{equation}
Note that unhatted variables are obtained (equivalent to performing an $n$ null vector rotation) by substituting $\Psi_4\leftrightarrow \Psi_0$ and $\Psi_1 \leftrightarrow \Psi_3$.

Transforming $\hat G$ under a Type I rotation will reproduce the sixth order equation for $\bar a$ in Eq.~(\ref{sixor}). This means that the polynomial can be written in terms of $\hat\alpha$, $\hat\beta$ and $\hat\gamma$. Hence,
\begin{equation}
\frac{\hat \alpha^2 \hat \beta^2 \hat \gamma^2}{16} = \hat G^2 \,,
\end{equation}
which increases the order of the polynomial from 6 to 12. However, it is now written as the product of three quartic equations. One of these equations, $\hat \alpha^2$, is associated with the principal eigenvalue $\lambda_{P}$, and the QK frame.

Under a Type I rotation, $\hat\alpha^2$ has the form
\begin{equation}
\label{alp2}
\hat\alpha^2 = B_{(1)}\bar a^4 + B_{(2)} \bar a^3 +B_{(3)} \bar a^2 + B_{(4)} \bar a+B_{(5)} \,,
\end{equation}
with coefficients
\begin{align}
B_{(1)} &=\lambda_1 \Psi_0 + 2 \Psi_1^2 - 2\Psi_0\Psi_2 \,,\cr
B_{(2)} &=4\lambda_1 \Psi_1 + \Psi_1\Psi_2 - \Psi_0 \Psi_3 \,,\cr
B_{(3)} &=6 \lambda_1\Psi_2 + 6 \Psi_2^2 - 4 \Psi_1\Psi_3 - 2\Psi_0\Psi_4 \,,\cr
B_{(4)} &=\lambda_1 \Psi_3 + \Psi_2\Psi_3 - \Psi_1 \Psi_4 \,,\cr
B_{(5)} &=2\Psi_3^2 + \Psi_4(\lambda_1 -2 \Psi_2) \,.
\label{quarticcoeffs}
\end{align} 
Instead of using this quartic directly, the authors go on to use a reduced variable 
\begin{equation}\label{ztrans}
    z=\Psi_0\bar a + \Psi_1 \,,
\end{equation}
to obtain a quartic
\begin{equation}\label{eq:quartic}
    Q_{(1)}z^4+Q_{(2)}z^3+ Q_{(3)}z^2 + Q_{(4)}z + Q_{(5)} =0 \,,
\end{equation}
whose coefficients are 
\begin{align}
    Q_{(1)} &=1 \,,\cr
    Q_{(2)} &=\dfrac{-4G}{\lambda_1 \Psi_0 -2H} \,,\cr
    Q_{(3)} &=\dfrac{6\Psi_0\lambda_1H + 6H^3 - 2K}{\lambda_1 \Psi_0 -2H} \,,\cr
    Q_{(4)} &=\dfrac{4G(H+\Psi_0\lambda_1)}{\lambda_1 \Psi_0 -2H} \,,\cr
    Q_{(5)} &=\dfrac{-2KH+2G^2 +\Psi_0\lambda_1 K}{\lambda_1 \Psi_0 -2H} \,,
\label{qcoeffs}
\end{align}
where 
\begin{align}
     G &= \Psi_0^2\Psi_3 - 3\Psi_0\Psi_1\Psi_2 + 2 \Psi_1^3 \,,\cr
     H&=\Psi_0\Psi_2 -\Psi_1^2 \,,\cr
     K&=\Psi_0^2 I -3H^2 \,,
\label{ghk}
\end{align}
which successfully reduces the sixth-order equation to fourth order.
In Ref.~\cite{Nerozzi:2004wv}, it is shown that the polynomial square root of this quartic equation can be found, reducing the order of the polynomial Eq.~(\ref{alp2}) to quadratic. 
 
From here, our methodologies will diverge from those in Ref.~\cite{Nerozzi:2004wv}, since we will deal with the quartic equation~(\ref{alp2}) that must first be solved and the correct root must then be appropriately chosen. Solving the depressed quartic equation~(\ref{dquar}) can be done analytically or numerically. In the following sections, the solution is found via Mathematica's built in \textit{Solve} function. While testing, we also employed an analytic method to reduce the quartic to a cubic equation, solved for the roots of the new cubic, and used them to construct the roots of the quartic. This is a standard analytic technique to find quartic equation solutions. Once the roots of the quartic are found, they are used in conjunction with Eqs.~(\ref{eq:TypeI}) and (\ref{eq:TypeII}) to rotate the Weyl scalars into the QK frame. 
 
In the following sections, we will employ both the QK frame, as well as the other transverse frames corresponding to the roots of $\hat\beta^2$ and $\hat\gamma^2$ or, equivalently and more simply, the eigenvalues not associated with the QK frame, in order to classify the strong-field region of an analytic spacetime.

\section{Binary black hole merger (approximate) local spacetime classification}\label{sec:Class}

The goal of this project is twofold: first, we would like to be able to use the QK frame to more accurately extract gravitational waves from a BBH system. Second, we would like to be able to locally classify an arbitrary black hole spacetime into its approximate Petrov type. The speciality invariant $S$ that we studied in the earlier part of this paper was introduced as a way to measure distortions from a Kerr (Petrov type D) spacetime~\cite{Baker:2000zm}. However, this invariant does not differentiate between Petrov types D and II. For $r\to\infty$ it is known that typical gravitational wave spacetimes are Petrov type D, but in the strong field region, we have seen that there are regions of algebraic speciality. Here, we introduce an index $\mathcal{D}$ which, if used in conjunction with $S$, can differentiate between a Petrov type II spacetime and a Petrov type D spacetime at points of algebraic speciality.
\begin{equation}\label{dindex}\centering
\mathcal{D}=\sqrt{\frac{12}{I}}\left(\Psi_2 - \frac{\Psi_3^2}{\Psi_4}\right) \,.
\end{equation}
To derive $\mathcal{D}$, return to Eq.~(\ref{ghk}).
The Petrov types can be characterized in terms of these scalars as shown in Table~\ref{classtable}. They correspond to the 
flow diagram for determining the Petrov type in Figure~9 of Ref.~\cite{Stephani:2003tm}.
\begin{table}[t]
\caption{Special Petrov types in terms of $G,\,H$ and $K$.}
\centering
\begin{tabular}{|l|l|}\hline
Petrov type & Characteristics\\\hline 
II & $G\neq 0$, $K-9H^2\neq 0$\\
D & $G=0$, $K-9H^2=0$, $K\neq0$\\
III, N, O &$J=0$, $I=0$\\
N & $G=0$, $H=0$\\
O & $G=0$, $K=0$, $H=0$\\
\hline
\end{tabular}

\label{classtable}
\end{table}
The $\mathcal{D}$ index can be derived from the condition in the second row of Table~\ref{classtable}, under the transformation, $\Psi_0\leftrightarrow\Psi_4$ and $\Psi_1\leftrightarrow\Psi_3$;
\begin{align}\label{dalitycondition}
    &K-9H^2=0 \,, \nonumber\\
    & \rightarrow \Psi_4^2I-12H^2 =0 \,, \nonumber\\
    & \rightarrow 1-\dfrac{12H^2}{\Psi_4^2I} = 1-\dfrac{12}{I}\left(\Psi_2 - \dfrac{\Psi_3^2}{\Psi_4} \right)^2=0 \,,\nonumber\\
    & \rightarrow 1-\mathcal{D}^2 = 0 \,.
\end{align}
This means that when $\mathcal{D}\to\pm1$ in an arbitrary frame that is not QK, the spacetime heads to either Petrov type D 
and this index will allow us to differentiate between Petrov types II and D for a region of algebraic speciality. 

In a QK frame, $\mathcal{D}_{\rm QK} = \pm 2$ in a Petrov type D spacetime. This can be proven as follows 
\begin{align}\label{eq:D_QK_2}\centering
\mathcal{D}_{\rm QK}&=\sqrt{\frac{12}{I}}\left(\Psi_2'' - \frac{\Psi_3''^2}{\Psi_4''}\right)\nonumber\\
&=\sqrt{\frac{12}{3\Psi_2''^2}}\,\Psi_2'' =\pm2 \,, 
\end{align}
whereas for an arbitrary transverse frame that is not QK, $\mathcal{D}=\pm1$ if a spacetime is Petrov type D.
That this is in general the case can also be seen from the relation $\mathcal{D}^2=4(1-\xi/I)$ in any transverse frame. In a QK frame we have $\xi\to0$ while in a non-QK frame $\xi\to3I/4$ in Petrov type D spacetimes.

For a spacetime in which $\Psi_4=0$, the equivalent version of the $\mathcal{D}$ index, with $\Psi_0\neq0$ can be obtained under the transformations $\Psi_1\leftrightarrow\Psi_3$ and $\Psi_0\leftrightarrow\Psi_4$.
A symmetrized version can also be obtained by adding the (averaged) corresponding $H,G$ and $K$ with $\Psi_4\to\Psi_0$ and $\Psi_3\to\Psi_1$ terms. Other expressions can also be obtained by including multiples of $G$ since it is vanishing for a Petrov type D spacetime. Those expressions will be equivalent once we choose a transverse frame.
Note that while $\mathcal{D}$ is invariant for Petrov type D spacetimes, for generic spacetimes it is only invariant under Type II and Type III (boost) tetrad rotations, hence it is frame dependent. That is why we will chose first a transverse frame to analyze the (approximate) classification of the spacetime.


\subsection{Analysis for Brill-Lindquist Initial Data} \label{BLResults}

In order to (1) verify that we are able to successfully rotate a spacetime into a QK frame and then (2) classify the strong-field region into different Petrov types, we have constructed a series of initial data tests on analytic systems of BBH pairs.
The first tests we performed of this rotation use analytic Brill-Lindquist initial data on an equal-mass binary system
with total mass $m=m_1+m_2$. The black holes are located at $z/m=\pm2.5,\,\cdots,\,\pm7.5$, and start from rest, so have separations $d/m=5,\,\cdots,\,15$ in increments of $1$. The system we will look at in-depth has $d/m=10$ and $z/m=\pm5$. 

We also studied a system with mass ratio $q=m_1/m_2=1/3$. In this system, the large black hole is located at $z/m=1.75$ and the small black hole is at $z/m=-5.25$ with masses $m_2/m=0.75$ and $m_1/m=0.25$ respectively. We will begin by showing in-depth results of the equal mass binary and then will move on to the unequal mass case. For this study, we will use the non-QK frame for classification. The lapse and shift we use are $N(r,\theta)$ and $\beta^i=0$.

The explicit values of the Weyl scalars in spherical polar coordinates are
\begin{widetext}
\begin{align}
\Psi_0 =& \dfrac{1}{r^2N^4\psi^6}
\left\{-\psi^2N,_\theta^2 + N \psi \left[-2 N,_\theta\psi,_\theta + \psi(-\cot\theta N,_\theta +N,_{\theta\theta}) \right] 
+N^2 \left[ 3\psi,_\theta^2 + \psi(\cot\theta\psi,_\theta - \psi,_{\theta\theta}) \right] \right\} \,, \\ 
\Psi_1 =& \frac{1}{2 \sqrt{2}r^2N^2\psi^6} \left\{-r\psi^2N,_\theta N,_r + N \psi\left[2 r (\psi,_\theta N,_r+ 2N,_\theta\psi,_r)+\psi (2 N,_\theta - r N,_{r\theta})\right] \right. \cr 
& \left. - 2N^2\left[3r\psi,_\theta\psi,_r + \psi(\psi,_\theta - r\psi,_{r\theta})\right] \right\} \,, \\
\Psi_2 =& \frac{1}{6r^2N^2\psi^6}\left\{-2\psi^2N,_\theta^2 + 4 N \psi N,_\theta\psi,_\theta + N^2 \left[ -3 \psi,_\theta^2 + \psi(\cot\theta \psi,_\theta +\psi,_{\theta\theta}) + \psi^2(-1+r^2N,_r^2)\right] \right. \cr
& \left. + r N^3\psi\left[-6r N,_r\psi,_r + \psi(-2 N,_r + r N,_{rr})\right] + N^4 \left[\psi^2 + 6r^2\psi,_r^2 + 2r\psi(\psi,_r-r\psi,_{rr})\right] \right\} \,, \\  
\Psi_3 =& \frac{1}{4\sqrt{2}r^2\psi^6} \left\{r\psi^2N,_\theta N,_r + N\psi\left[-2r(\psi,_\theta N,_r + 2 N,_\theta \psi,_r) + \psi(-2 N,_\theta + r N,_{r\theta})\right] \right. \cr
& \left. + 2N^2 \left[3 r\psi,_\theta\psi,_r + \psi(\psi,_\theta - r \psi,_{r\theta})\right] \right\} \,, \\
\Psi_4 =& \frac{1}{4r^2\psi^6} \left\{-\psi^2N,_\theta^2 + N \psi\left[-2N,_\theta\psi,_\theta+ \psi(-cot\theta N,_\theta + N,_{\theta\theta})\right] + N^2 \left[3\psi,_\theta^2 + \psi(\cot\theta\psi,_\theta - \psi,_{\theta\theta})\right] \right\} \,,
\end{align}
\end{widetext}
with conformal factor $\psi$. 
They are constructed using the analytic tetrad in Brill-Lindquist Coordinates~\cite{Lousto:1999bd} 
\begin{align}
l^\mu_{\rm an} &= \left\{ \frac{1}{N^2},\,\frac{1}{\psi^{2}},\, 
0,\, 0 \right\} \,,
\cr
n^\mu_{\rm an} &= \frac{1}{2} \left\{ 1,\, 
- \frac{N^2}{\psi^{2}},\, 0,\, 
0\right\} \,, 
\cr
m^\mu_{\rm an} &= \frac{1}{\sqrt{2}r\psi^{2}} \left\{ 0,\, 0,\, 1,\,  
\frac{i}{\sin\theta}\right\} \,, 
\label{eqn:num_tetrad}
\end{align}
and then rotated into a transverse frame using the process in Sections~\ref{sec:qkintro}--\ref{QKrotation}.
In what follows, for the sake of simplicity, as in Ref.~\cite{Lousto:1999bd}, we will chose
$-N^2=-(1-2m/r)$.

\subsubsection{Equal mass $d/m=10$ case}
\label{q1}

\begin{figure}[t]\centering
	\includegraphics[width=0.99\columnwidth]{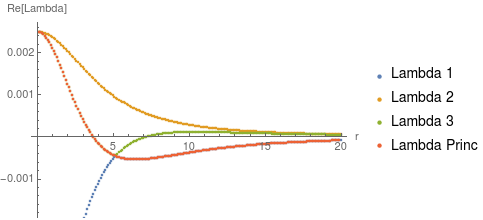}
	\caption{Eigenvalues for a BBH system with Brill-Lindquist initial data for $q=1$ on the symmetry plane $\theta=\pi/2$. The black holes are at $r/m=5$ on the $z$-axis, i.e., $z/m=\pm5$. At approximately $r/m=5$, the principal eigenvalue must be switched from $\lambda_1$ so that it remains smooth.}
	\label{lambdapfig}
\end{figure}

Figure~\ref{lambdapfig} shows all three choices for the eigenvalues, $\lambda_1,\,\lambda_2$ and $\lambda_3$, as well as highlights the principal eigenvalue $\lambda_P$ (in red) that is associated with the QK frame. These correspond to analytic initial data for a Brill-Lindquist binary with separation $d/m=10$ and equal masses versus $r$, and are shown on the symmetry plane $\theta=\pi/2$. 
Notice first that at $r/m=5$ (which we will henceforth refer to as $r=r_{\rm Ring}$) $\lambda_P$ changes branches from $\lambda_1$ to $\lambda_3$.
This is done because we can, in fact, demand continuity from our principal eigenvalue~\cite{Beetle:2004wu}. The eigenvalue branch that is QK only need be switched when $r=r_{\rm Ring}$, not necessarily whenever $S=1$, which is true not only at $r=r_{\rm Ring}$, but also asymptotically and between the black holes.

Notice that one could choose $\lambda_2$ to be principal everywhere since it is continuous, use this to do the rotation, and transform the Weyl scalars into a transverse frame. Since this eigenvalue does not satisfy $\lambda_{P} =\text{max}_k|\lambda_k|$ as $r\to\infty$, the frame that uses $\lambda_2$ as principal is transverse, but is not QK and therefore asymptotically may not lead to tetrad vectors that are near Kinnersley. It will, however, be of interest to us later on to use an alternative transverse frames to classify the spacetime into different Petrov types.

\begin{figure}[t]\centering
	\includegraphics[width=0.99\columnwidth]{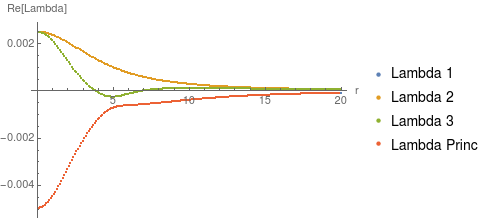}
	\caption{Eigenvalues for a BBH system with Brill-Lindquist initial data for $q=1$ at $\theta=8\pi/15$. The black holes are at $z/m=\pm5$. On the whole spacetime, the principal eigenvalue remains $\lambda_1$.}
	\label{lambdapfig2}
\end{figure}

When we look just off the symmetry plane (for example, if $\theta=8\pi/15$), the principal eigenvalue remains $\lambda_P=\lambda_1$ for the whole spacetime, which means that the surface $r=r_{\rm Ring}$ is one dimensional in shape. Figure~\ref{lambdapfig2} shows this case, with the principal branch in red. We can clearly see that all three eigenvalues, and thus transverse frames, are continuous even in the strong-field region.
This means that the points at $(r_{\rm Ring},\,\theta_{\rm Ring})=(r_{\rm Ring},\,\pi/2)$ are the only locations where the eigenvalues have a cusp when $S=1$. They form a one dimensional ring of points, that later we will be shown to be encompassed by a three dimensional doughnut shape. This will be studied more in Section~\ref{ds0}.

On the symmetry plane $\theta=\pi/2$, we have found a principal eigenvalue $\lambda_P$ that is continuous throughout the whole spacetime and can now insert it into Eq.~(\ref{qcoeffs}) to construct the quartic equation~(\ref{alp2}).
Solving any quartic gives at most four roots $a_1,\,a_2,\,a_3$ and $a_4$. Of these four solutions, only two are associated with the QK frame (two instead of one due to $l\leftrightarrow n$ degeneracy). 

To determine which two roots we want, we will introduce the radiation scalar $\xi^{\rm QK}= \Psi_0^{\rm QK}\Psi_4^{\rm QK}$ in the QK frame which is used to classify the spacetime in Ref.~\cite{Nerozzi:2004wv} in the far-field region. However, no classification criteria beyond continuity is provided for the strong-field region. Since in the QK frame, both $\Psi_4^{\rm QK}$ and $\Psi_0^{\rm QK} \to 0$ when $S\to1$, it must be true that $\xi^{\rm QK}\to0$ as well. This will be useful when choosing the correct root from $a_i$. 

\begin{figure}[t]\centering
	\includegraphics[width=0.99\columnwidth]{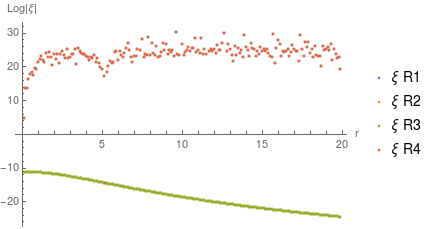}
	\caption{A plot of $\log|\xi|$ ($\xi=\Psi_0 \Psi_4$ where we set $m=1$) for $\Psi_0$ and $\Psi_4$ in all four transverse frames on the plane $\theta=\pi/2$. The scalars $\xi_2$ and $\xi_3$ are QK since $\xi\to0$ as $S\to 1$ (equivalently, $r\to\infty$). The red point in the green curve comes from the fact that there is only one transverse frame at that point, so all four values of $\xi$ coincide.}\label{xifig}
\end{figure}

Figure~\ref{xifig} shows $\log |\xi|$ (where we set $m=1$) on the equatorial plane for all four transverse frames associated with roots $a_1,\,a_2,\,a_3$ and $a_4$. The values of $\xi$ for roots $a_1$ and $a_4$ and roots $a_2$ and $a_3$ (respectively) coincide. Since $\xi_2$ and $\xi_3$ head to 0 as $r\to\infty$ and are both continuous, the roots associated with the QK frame are $a_2$ and $a_3$. Interestingly, at $r=r_{\rm Ring}$, where we switch which eigenvalue branch is designated principal, $\xi_1$ and $\xi_4$ coincide with $\xi_2$ and $\xi_3$ (hence the red point at $r=r_{\rm Ring}$ located at about $\xi=10^{-15}$ on the green curve). This supports the claim that at $r=r_{\rm Ring}$ only one transverse frame exists, since all roots produce the same value of the scalar $\xi$ at this point (and we will later show that this is characteristic of Petrov type II spacetimes).

The disadvantage to this method of root classification is that $\xi$ must be computed for all four roots at each point in the entire spacetime (not just on the symmetry plane, as is shown in Figure~\ref{xifig}). This requires both Type I and Type II rotations of both $\Psi_0$ and $\Psi_4$, and at least a Type I rotation of all other Weyl scalars. For a small-scale analytic calculation, this is not a problem. For a large scale numerical BBH simulation where this must be done at all points on a 3D grid at every timestep, it could become computationally inefficient. 

\begin{figure}[t]\centering
	\begin{minipage}{0.45\textwidth}
		\centering
		\includegraphics[width=0.99\columnwidth]{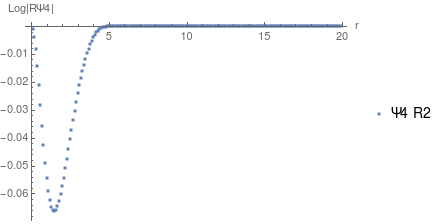}
	\end{minipage}
	\begin{minipage}{0.45\textwidth}
		\centering
		\includegraphics[width=0.99\columnwidth]{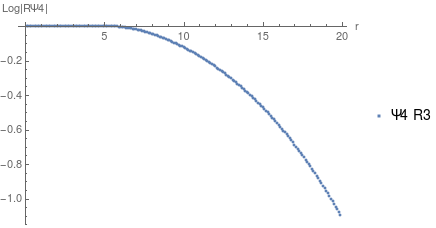}
	\end{minipage}
	\caption{The value of $\Psi_4$ (where we set $m=1$) for the two roots $a_2$ and $a_3$ (top and bottom), where $\xi\to 0$ as $S\to1$. For the root $a_2$, $\Psi_4$ has the expected behavior, heading to 0 as $r\to\infty$, for the root $a_3$, $\Psi_4$ diverges.}
	\label{p423}
\end{figure}

Even though there are two ``correct'' choices for the root associated with the QK frame, it may be true that $\Psi_0$ or $\Psi_4$ do not tend to 0 as $r\to\infty$. 
It follows that in order to determine which of the two transverse frames associated with $a_2$ and $a_3$ is QK, one can look at the values of $\Psi_0$ and $\Psi_4$ individually as $r\to\infty$.
The QK frame is found by choosing the root that minimizes the magnitude of both $\Psi_0$ and $\Psi_4$ for large $r$.  Both $a_2$ and $a_3$ produce $\Psi_0$ that are continuous and head to 0 as $r\to\infty$. However, there are differences in asymptotic behavior in $\Psi_4$: using $a_2$, $\Psi_4$ is continuous and heads to 0 as $r\to\infty$, but using $a_3$, $\Psi_4$ grows exponentially as $r\to \infty$.
Figure~\ref{p423} shows the real parts of $\Psi_4$ calculated using the roots $a_2$ and $a_3$, respectively. 
Once the correct root is chosen, we can compute the Weyl scalars in the QK frame. 

The implementation of this method is not so straightforward. In addition to the choices that need to be made asymptotically, we are attempting to demand continuity in the strong field region as well. This is subject to complex number arithmetic issues, as well as branch changes among other practical difficulties, especially for binaries that are orbiting or spinning. This is why we have chosen to begin with only analytic results for head on collision configurations. Future work should be done to extend these analyses to the strong-field region of more complicated systems in order to use it for extraction of gravitational waves.


\subsubsection{Results: Characterization of maxima, minima, and zeros of $S$ and $\mathcal{D}$}\label{ds1}

The goal of our work is to use $\mathcal{D}$ in conjunction with $S$ to do a point-by-point analysis of the approximate Petrov type of a spacetime, with a specific focus on the strong field region and between the black holes where there is no a priori knowledge of the spacetime's Petrov type. For this work, consider again the $q=1$, $d/m=10$ Brill-Lindquist initial data binary.

\begin{figure}[t]\centering
	\begin{minipage}{0.45\textwidth}
		\centering
		\includegraphics[width=0.99\columnwidth]{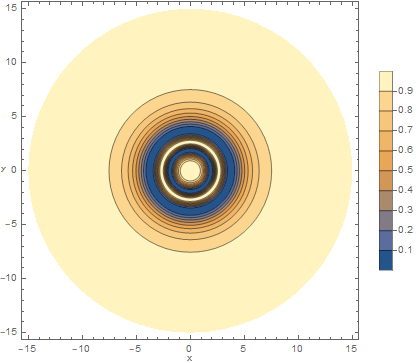}
	\end{minipage}
	\begin{minipage}{0.45\textwidth}
		\centering
		\includegraphics[width=0.99\columnwidth]{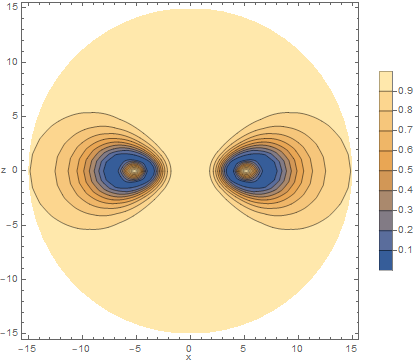}
	\end{minipage}
	\caption{Speciality invariant $S$ on the $xy$- and $xz$-planes (top/bottom) for a $q=1$ binary with separation $d/m=10$ located on the $z$-axis. Other than far from the black holes and between the black holes, the only region of algebraic speciality is located in a ring on the $x$-axis at $r/m=5$.}
	\label{fig:SxySxz_q1}
\end{figure}

To begin, look at the invariant $S$ on the $xy$- and $xz$-planes (see the top and bottom panels of Figure~\ref{fig:SxySxz_q1}, respectively). Recall that when $S=1$, the spacetime is algebraically special. 
As $r\to\infty$ we expect the spacetime to be Petrov type D and therefore algebraically special, and in fact, in this region, $S\to1$ in both the top and bottom panels of Figure~\ref{fig:SxySxz_q1}. 
The $xy$-plane exhibits algebraic speciality between the black holes in addition to when $r\to\infty$. Interestingly, but not unexpectedly given our earlier analysis, there is a ring of algebraic speciality at $r=r_{\rm Ring}$. This is visible in the $xz$-plane (the bottom panel of Figure~\ref{fig:SxySxz_q1}) as well --- there are two points at $x=\pm r_{\rm Ring}$ where $S=1$. In fact, on the $xz$-plane, $S\to1$ everywhere except in ellipsoidal regions surrounding the points where $r=r_{\rm Ring}$. 

Where $S\to1$, it is known the spacetime is either Petrov type II or D~\cite{Baker:2000zm}. Close to and between the black holes, $S\to1$ does not characterize the Petrov type of the points of the spacetime since it cannot differentiate between Petrov types II and D where there is no a priori knowledge of the spacetime's behavior. To remedy this, we propose using the $\mathcal{D}$ index from Eq.~(\ref{dindex}) to provide an approximate Petrov characterization of the points in the strong-field region.
 
\begin{figure}[t]\centering
	\includegraphics[width=0.99\columnwidth]{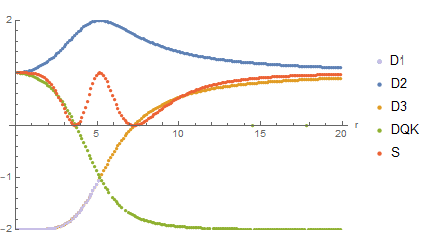}
	\caption{$\mathcal{D}$-index and $S$-invariant in the QK frame for the $q=1$, $d/m=10$, BBH case with Brill-Lindquist initial data on a slice through the symmetry plane $\theta=\pi/2$. The index $\mathcal{D}$ associated with the QK frame goes to -2 asymptotically, intersects with $\mathcal{D}_3$ and then switches branches to be on the branch associated with $\lambda_3$. 
	}
	\label{dns}
\end{figure}

Recall Figure~\ref{lambdapfig}; which shows the eigenvalues of the matrix $Q_{ab}$ for this system on a slice through the equatorial plane at time $t=0$. The eigenvalues $\lambda_1$ and $\lambda_3$ have cusps when $S=1$ at $r=r_{\rm Ring}$. This means that the only eigenvalue that exists on this ring of algebraic speciality is $\lambda_2$. This is particularly interesting; it implies that only one transverse frame actually exists here and this frame must \textit{not} be QK since $\lambda_2$ is not associated with the QK frame. In fact, according to Appendix C of Ref.~\cite{Nerozzi:2004wv}, a spacetime with exactly one transverse frame must be Petrov type II~\footnote{In Appendix C.2 of Ref.~\cite{Nerozzi:2004wv}, there is a typo:  $\Psi_0''\Psi_4''=(9/4)\,(\Psi_2)^2$ instead of $3/2$. This makes $\mathcal{D}=1$ for a Petrov type D spacetime.} and infinitely many transverse frames must be Petrov type D.
This means that we can use $\mathcal{D}_{\rm QK}$ for classification in the far-field region, but at $r=r_{\rm Ring}$ we cannot. The only viable transverse frame, the one that is continuous over the whole spacetime, is associated with $\lambda_2$, which is consistent with the frame being Petrov type II at $r=r_{\rm Ring}$. We will call the associated classification index $\mathcal{D}_2$.
Figure~\ref{dns} shows the corresponding $\mathcal{D}$ index in all three transverse frames --- the QK frame as well as the other two non-QK transverse frames, and the $S$-invariant.

At $r=r_{\rm Ring}$, we have $\mathcal{D}_2=2$. This means that the points of the spacetime on $r=r_{\rm Ring}$ should be Petrov type II, and we can prove this as follows.
We have already seen that $\Psi_0''\Psi_4''\to 0$, and since our frame is transverse, $\Psi_3''=0$ as well. So,
$\mathcal{D}=\pm2$.
Note that $\xi$ vanishes by Table I of Ref.~\cite{Nerozzi:2004wv} in a Petrov type II spacetime.

The QK frame definition forces continuity through $r=r_{\rm Ring}$ in $\lambda_{P}$ by setting $\lambda_{P}=\lambda_3$ when $r<r_{\rm Ring}$ instead of continuing on the branch $\lambda_P=\lambda_1$.
Therefore,  and between the black holes, where $r=0$, $\mathcal{D}=S=1$ implies that the spacetime heads to Petrov type D. Because this switching between frames is done, we would need to shift how we characterize the spacetime when we move past $r=r_{\rm Ring}$ which can easily lead to classification errors, so it is best to use a branch that is natively continuous throughout the whole spacetime. 
In our Brill-Lindquist system, we have already determined that is the one associated with $\lambda_2$. We can therefore look to $\mathcal{D}_2$ which, as $r\to0$, heads to 1. This is verification that the Petrov type between the black holes approaches D.  

\subsubsection{Classification of the region where $S=0$}
\label{ds0}

\begin{figure}[t]\centering
	\begin{minipage}{0.45\textwidth}
		\centering
		\includegraphics[width=0.9\columnwidth]{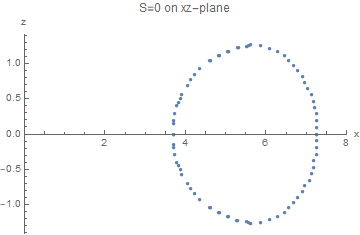}
	\end{minipage}
	\begin{minipage}{0.45\textwidth}
		\centering
		\includegraphics[width=0.9\columnwidth]{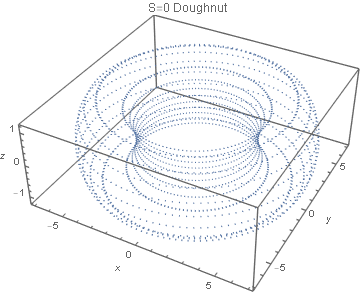}
	\end{minipage}
	\caption{The ring of points on the $xz$-plane where $S=0$ (top). This ring rotates around the $z$-axis, forming a hollow ``doughnut'' of points in space where $S=0$ (bottom). As resolution increases, the points will approach a continuous ring on the $xz$-plane.}
	\label{fig:Sdonut}
\end{figure}

In Figure~\ref{dns}, at approximately $r/m=3.68$ and $7.28$, $\mathcal{D}=S=0$. In a spacetime where $S=J=0$, it can be shown that $\mathcal{D}=\pm\sqrt{3} \text{ or } 0$:
\begin{align*}
    \mathcal{D}&=\sqrt{\frac{12}{I}} \Psi_2''\\
    &= \mathrm{sgn}(\Psi_2'')\,\sqrt{\frac{12\Psi_2''^2}{3\Psi_2''^2+\Psi_0''\Psi_4''}} \,,
\end{align*}
where $J=0$ implies 
\begin{equation*}
    \Psi_0''\Psi_4''=\Psi_2''^2 ~~\text{or}~~ \Psi_2''=0 \,.
\end{equation*}
Therefore, we have
\begin{equation*}    
    \mathcal{D}_{(S=0)}=\pm\sqrt{3} ~~\text{or}~~ 0 \,,
\end{equation*} 
where $\mathcal{D}=0$ if $\Psi_2''=0$. 
Recall again the plots in Figure~\ref{fig:SxySxz_q1}. In both the $xy$- and $xz$-planes, there exist regions of the spacetime where $S\to0$. Unlike the $S\to1$ ring, $S\to0$ at two points on $\theta=\pi/2$ ($z=0$), one to the left of $r=r_{\rm Ring}$ at $r/m=3.68$ and one to the right at $r/m=7.28$. In fact, $S=0$ on a ring on the $xz$-plane itself. The top panel of Figure~\ref{fig:Sdonut} shows this ring in Quadrants I and IV of the $xz$-plane, but the $\phi$-symmetry of this system means the ring rotates around the $z$-axis to form a hollow ``doughnut"-shape (the bottom panel of Figure~\ref{fig:Sdonut}).
The $S=0$ doughnut's outer and inner rings have radii of $r/m=7.28$ and $3.68$ from the origin, respectively, along the $x$-axis. The doughnut has $z-$maxima at $(x/m,\,z/m)=(\pm5.90,\,1.25)$ and minima at $(x/m,\,z/m)=(\pm5.90,\,-1.25)$. This region of spacetime surrounds the surface of algebraic speciality located at $r_{\rm Ring}/m \approx 5.05$, but is not itself algebraically special since $S=0$, and is instead a general Petrov type I.

Since we have concluded that $r=0$ and $r\to\infty$ are Petrov type D and $r=r_{\rm Ring}$ is Petrov type II, we will argue that on this doughnut where $S=0$, the spacetime is transitioning between Petrov types II and D. To investigate this region, we will use the $\mathcal{D}$ index in the transverse frame associated with $\lambda_2$ because it is continuous through the point $r=r_{\rm Ring}$, rather than the QK frame. This ensures that we are not switching between frames, that all interesting points exist, and reduces the likelihood of classification error. Recall that this means that $\mathcal{D}=1$ in a Petrov type D spacetime. We have already shown that when $\mathcal{D}=\pm2$ the spacetime is Petrov type II in Eq.~(\ref{eq:D_QK_2}).

In Figure~\ref{dns}, notice that $\mathcal{D}_2$ is not symmetric around $r=r_{\rm Ring}$. Even so, on the left and right hand side at $x/m=3.68$ and $x/m=7.28$ respectively, $\mathcal{D}_2=1.73$. Interestingly, this means that the spacetime should be closer to Petrov type II than D at these points because a larger value of $\mathcal{D}_2$ implies a smaller value of $\Psi_0''\Psi_4''$ which leads to $\mathcal{D}_2\to-2$.

Table~\ref{tab:qksummq1} shows the $r$-locations of important values on $\theta=\pi/2$. Namely, the location of the ring, 
and the values of $\mathcal{D}$ and $S$ at important points in different transverse frames (QK, 2, or 3).
\begin{table}[t]
    \caption{Summary of the values of the scalars $S$ and $\mathcal{D}$ at different $r$-locations in different Transverse frames (QK, 1, 2, and 3) and the associated Petrov type for the $q=1$ Brill-Lindquist binary on $\theta=\pi/2$. We are using the transverse frame 2 
    for classification. When $\mathcal{D}=\pm1.73$ and $S=0$ the spacetime is Petrov type I, but is closer to Petrov type II than Petrov type D. When $\mathcal{D}=\pm1.5$, the spacetime is halfway between Petrov types II and D. 
    }
    \centering
    \begin{tabular}{|c|c|c|c|c|}\hline
         Transverse Frame& $S$&$\mathcal{D}$ &$r/m$&Petrov type \\\hline\hline
         &1&-2&0&D\\
         &0.5&0.52&2.68&I\\
         &0&0&3.8&I\\
         1&0.5&-1.41&4.9&I\\
         &1&-1&5.05&II\\
         QK $(1\to3)$ &0.5&-1.41&5.1&I\\
         &0.32&-1.5&6.27&I\\
         &0&-1.73&7.20&I\\
         &0.5&-1.93&9.88&I\\
         &1&-2&$\infty$&D\\
         \hline
         &1&1&0&D\\
         &0.5&1.41&2.68&I\\
         &0.31&1.5&2.97&I\\
         &0&1.73&3.8&I\\
         &0.5&1.91&4.9&I\\
         2&1&2&5.05&II\\
         &0.5&1.91&5.1&I\\
         &0&1.73&7.2&I\\
         &0.31&1.5&8.97&I\\
         &0.5&1.41&9.88&I\\
         &1&1&$\infty$&D\\
         \hline 
         &1&1&0&D\\
         &0.5&-1.93&2.68&I\\
         &0&-1.73&3.80&I\\
         &0.32&-1.5&4.26&I\\
         &0.5&-.42&4.9&I\\
         3&1&-1&5.05&II\\
         &0.5&-.42&5.1&I\\
         &0&0&7.2&I\\
         &0.5&0.52&9.88&I\\
         &1&1&$\infty$&D\\
    \hline         
    \end{tabular}
    \label{tab:qksummq1}
\end{table}

\subsubsection{Classification off the symmetry plane}
\label{class8pi15}
\begin{figure}[t]\centering
	\includegraphics[width=0.99\columnwidth]{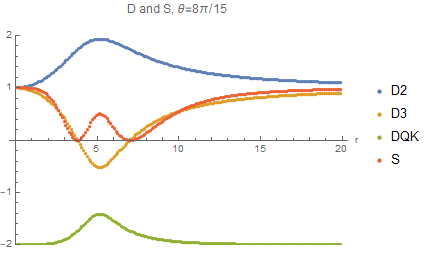}
	\caption{$\mathcal{D}$-index and $S$-invariant in the QK frame for the $q=1$, $d/m=10$, BBH case with Brill-Lindquist initial data on a slice through cone $\theta=8\pi/15$.}
	\label{dns8pi15}
\end{figure}

If we move off of the symmetry plane $\theta=\pi/2$ to a neighboring region, say the cone $\theta=8\pi/15$ to be consistent with Figure~\ref{lambdapfig2}, we can use any of the transverse frames for classification since they are all continuous. Figure~\ref{dns8pi15} shows the corresponding values of $\mathcal{D}$ and S for the transverse frames associated with the cone $\theta=8\pi/15$.
As $r\to0$ and $r\to\infty$, the two transverse, but not QK, frames, $\mathcal{D}_2$ and $\mathcal{D}_3$ go to 1, whereas $\mathcal{D}_{\rm QK}\to-2$. This is consistent with Petrov type D behavior in these regions. At $r=r_{\rm Ring}$, the values of $\mathcal{D}$ do not quite reach 2 (for non-QK) and 1 (for QK), and therefore the spacetime is never exactly Petrov type II, but only Petrov type I in the region between the points where $S=0$. This is consistent with our findings since (1) $S\neq1$ in this region, so there is no point of algebraic speciality, and (2) all three transverse frames exist --- that cannot be the case in a Petrov type II spacetime.

From these studies, we can draw the conclusion that between the points where $S=0$ (namely, around $r=r_{\rm Ring}$, the spacetime is actually closer to Petrov type II than it is to Petrov type D and is exactly Petrov type II only on the one dimensional ring at $\theta=\pi/2$.

\subsubsection{Unequal mass binary case}
\label{sec:q3resultsqk}
To generalize our results from Section~\ref{BLResults}, we studied a spacetime with an unequal mass binary whose mass ratio is $q=1/3$. This system also has analytic Brill-Lindquist initial data with $d/m=7$ separated black holes, and $z_1/m=-5.25$ and $z_2/m=1.75$ with the center of mass located on the origin of coordinates. The black holes have masses $3m_1/m=0.75=m_2/m$.
In particular, we are interested in (1) the value of $r_{\rm Ring}$, (2) the Petrov classification using the $\mathcal{D}$ index, and (3) the location of the doughnut $S=0$. 

Let us consider the system with Brill-Lindquist initial data with mass ratio $q=1/3$, and varying separation $d/m=7$. The black holes have masses $3m_1/m=0.75=m_2/m$ and are located at respective distances of $z_1=-5.25$ and $z_2=1.75$
($z_1=-d/(1+q)$ and $z_2=d q/(1+q)$ on the $z$-axis so that the center of mass is always located at the origin).
We are looking to find the one dimensional surface characterized by some $(r_{\rm Ring},\,\theta_{\rm Ring})$. We know that, if it exists, the circle occurs at some region where $|S-1|=0$. This will also be the location that, in order to maintain continuity in the eigenvalues $\lambda$ of the associated matrices $Q_{ab}$, the eigenvalue branch must be flipped. In our implementation, we used that as our criteria for selecting the appropriate $(r,\,\theta)$ pair to define the location of $(r_{\rm Ring},\theta_{\rm Ring})$. 

\begin{figure}[t]\centering
	\begin{minipage}{0.45\textwidth}
		\centering
		\includegraphics[width=0.99\columnwidth]{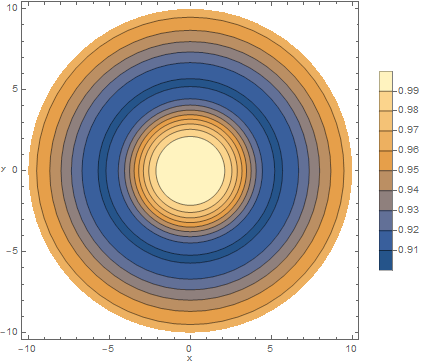}
	\end{minipage}
	\begin{minipage}{0.45\textwidth}
		\centering
		\includegraphics[width=0.99\columnwidth]{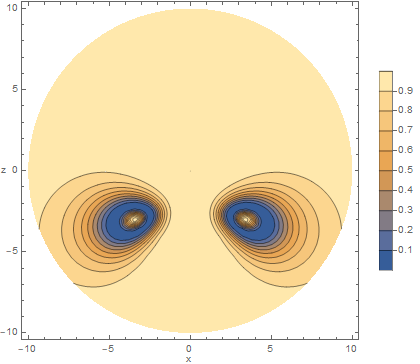}
	\end{minipage}
	\caption{Speciality invariant $S$ on the $xy$- and $xz$-planes (top/bottom) for a $q=1/3$ binary with separation $d/m=7$ located on the $z$-axis. Other than far from the black holes and between the black holes, the only region of algebraic speciality is located in a ring with radii in the centers of the ``eyes" in the bottom plot. The coordinates of the ring are $(r_{\rm Ring}/m,\,\theta_{\rm Ring})\approx(4.6,\,19\pi/15)$}
	\label{fig:SxySxz_q3}
\end{figure}

\begin{figure}[t]\centering
	\includegraphics[width=0.99\columnwidth]{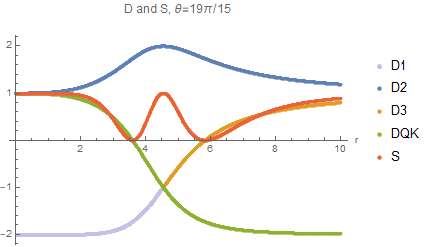}
	\caption{$\mathcal{D}$-index and $S$-invariant in the QK frame for the $q=1/3$, $d/m=7$, BBH case with Brill-Lindquist initial data on the conical slice $\theta=19\pi/15$. 
}
	\label{dnsq3}
\end{figure}

We found that the ring is located on a cone at $\theta_{\rm Ring}\approx19\pi/15$ regardless of binary separation $d$, so long as the origin is on the center of coordinates. The radius for the particular configuration with $d/m=7$ occurs at approximately $r_{\rm Ring}/m\approx4.9$, which can be seen from Figures~\ref{fig:SxySxz_q3} and~\ref{dnsq3}.

We would like to next generalize the classification results we saw in our study of the $q=1$ binary in Section~\ref{ds0} by extending that work to the case of the $q=1/3$ binary (and hence infer about other mass ratios). This, again, will be done with the use of the index $\mathcal{D}$ in conjunction with the Baker-Campanelli speciality invariant $S$. 
Figure~\ref{fig:SxySxz_q3} shows the $S$-invariant on the $xy$- and $xz$-planes in the top and bottom panels, respectively. Note the change in color scale between the two figures. The $xy$-plane, the top panel of Figure~\ref{fig:SxySxz_q3}, is symmetric on both axes due to $\phi$-symmetry, and has the property that $S=1$ both between the black holes as well as asymptotically. It drops to $S\approx0.9$ on a ring that corresponds to where the ``eyes'' cross the $x$-axis in the bottom panel of Figure~\ref{fig:SxySxz_q3}. On the $xz$-plane, the blue ellipsoidal regions are where $S=0$. When rotated around the $z$-axis they become ``doughnut-like'' in shape, which corresponds with what we saw in the bottom panel of Figure~\ref{fig:SxySxz_q1}. The value of $S$ at points asymptotically and between the black holes is 1, indicating a region of algebraic speciality over the whole spacetime except near the doughnut. There also exist points where $S=1$ inside of the doughnut at $(r_{\rm Ring}/m,\,\theta_{\rm Ring})\approx(4.9,\,19\pi/15)$ which, when rotated around the $z$-axis, forms the ring of algebraic speciality --- and consequently the location where two of the three eigenvalues $\lambda$ cease to exist. We expect this ring is Petrov type II, and seek to show that in what follows. 

Figure~\ref{dnsq3} shows the values of $\mathcal{D}$ in the QK frame as well as in the two other transverse frames overlayed with $S$. All are shown on the cone $\theta_{\rm Ring}\approx19\pi/15$. Analogously to Figure~\ref{dns}, $\mathcal{D}_3$ and $\mathcal{D}_{\rm QK}$ cross at $r=r_{\rm Ring}$, so forcing continuity means we have to switch frames when crossing $r=r_{\rm Ring}$. Furthermore, the frames themselves do not exist at the point of crossing, and only one transverse frame, associated with $\mathcal{D}_2$ exists and is equal to +2. This means that at the point $r=r_{\rm Ring}$ the spacetime is of Petrov type II, which is consistent with our results for the $q=1$ case. Far away from, as well as between the black holes, $\mathcal{D}_2=S=1$ which indicates that the spacetime is of Petrov type D in these regions.

Now that our classification results from the $q=1$ case are confirmed for a $q=1/3$ binary, we can look at the region that we expect is Petrov type I. The ``doughnut'' where $S=0$ exists in this unequal mass case, as well as in the equal mass binary. 
\begin{figure}[t]\centering
	\begin{minipage}{0.45\textwidth}
		\centering
		\includegraphics[width=0.9\columnwidth]{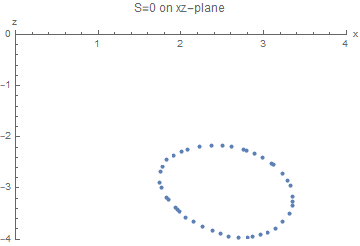}
	\end{minipage}
	\begin{minipage}{0.45\textwidth}
		\centering
		\includegraphics[width=0.9\columnwidth]{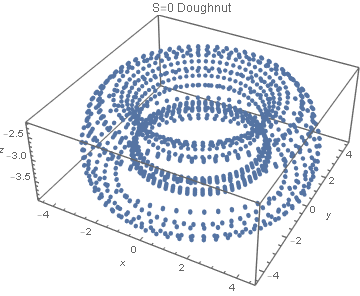}
	\end{minipage}
	\caption{The location of the doughnut $S=0$ on the $xz$-plane for the $q=1/3$, $d/m=7$ separated Brill-Lindquist system (top). Because of $\phi$-symmetry, the ellipsoid shown is rotated around the $z$-axis forming a hollow doughnut where $S=0$ (bottom). }
	\label{fig:donutq3}
\end{figure}
The ellipsoid shown in the top panel of Figure~\ref{fig:donutq3} outlines the region where $S=0$ on the $xz$-plane (the top panel) and, when rotated around the $z$-axis, will form the ``doughnut'' shape (the bottom panel). Interestingly, on the bounds of this region, $\mathcal{D}_2=-1.73$, are the same as in the equal mass case  (results shown in Table~\ref{tab:qksummq1}). This is a good indication that $\mathcal{D}$ and $S$ on $\theta=\theta_{\rm Ring}$ have a consistent relationship among different mass-ratios. We can hence safely say that the interior of the doughnut is approximately of Petrov type I, transitioning to Petrov type II at $r=r_{\rm Ring}$.

Table~\ref{tab:qksummq3} shows the $r$ locations of important values on the cone $\theta=19\pi/15$ --- namely the approximate location of the ring and the values of $\mathcal{D}$ and $S$ at important points in different transverse frames (QK, 2, and 3). 
\begin{table}[t]
    \caption{Summary of the values of the scalars $S$ and $\mathcal{D}$ at different $r$-locations in different Transverse frames (QK, 1, 2, and 3) and the associated Petrov type for the $q=1$ Brill-Lindquist binary on $\theta=19\pi/15$. Note that for $r\leq r_{\rm Ring}$, frames QK and 3 flip places when $S=1$ at $r=r_{\rm Ring}$.}    
    \centering
    \begin{tabular}{|c|c|c|c|c|}\hline
         Transverse Frame& $S$&$\mathcal{D}$ &$r/m$&Petrov type \\\hline\hline
         &1&-2&0&D\\
         1&0&-1.73&3.8&I\\
         &1&-1&4.9&II\\
         QK ($1\to3$)&0&-1.73&5.8&I\\
         &1&-2&$\infty$&D\\\hline
         &1&1&0&D\\
         &0&1.73&3.8&I\\
         2&1&2&4.8&II\\
         &0&1.73&5.8&I\\
         &1&1&$\infty$&D\\\hline 
         &1&1&0&D\\
         &0&0&3.8&I\\
         3&1&-1&4.9&II\\
         &0&0&5.8&I\\
         &1&1&$\infty$&D\\\hline \hline        
    \end{tabular}
    \label{tab:qksummq3}
\end{table}

\subsection{S-invariant Surface Levels}

In light of the previous example, we can turn now to a generic study of the values of the $\mathcal{D}$-index in the transverse frames given constant values of the invariant $S=\bar{\sigma}$. From its definition shown in Eq.~(\ref{eq:S}) and the expressions of Eqs.~(\ref{Iinv}) and~(\ref{Jinv}) for the invariants $I$ and $J$ in the transverse frame, we find
\beq\label{eq:sigma}
S=\bar{\sigma}=27\, \frac{(\bar{\xi}-1)^2}{(\bar{\xi}+3)^3} \,,
\eeq
where $\bar{\xi}=\Psi_{0}^{\rm TF}\Psi_{4}^{\rm TF}/(\Psi_{2}^{\rm TF})^2$ in the transverse frames.
In this notation, the index $\mathcal{D}$ takes the simple form
\beq\label{eq:D2}
\mathcal{D}^2=\frac{12}{(\bar{\xi}+3)} \,.
\eeq

Eq.~(\ref{eq:sigma}) is simple to solve numerically for a given value of $S=\bar{\sigma}$.
Table~\ref{tab:sigma} provides a few reference values for our analysis.
This is in complete agreement with our previous specific studies for Brill-Lindquist data
and provides a measure of deviations from the algebraic special case $S=1$ in terms
of the $\mathcal{D}$-index. When a given spacetime has the potential of being locally of 
Petrov type II, we then expect two of the eigenvalues to collapse and leave only room for 
a single well-behaved eigenvalue at which point evaluating the value of $\mathcal{D}$ in
that (non-QK) frame would be able to discriminate between Petrov types II 
and D.

\begin{table}[t]
    \caption{Values of $\mathcal{D}$ on the three transverse frames for a given value of $S=\bar{\sigma}$.}
    \centering
    \begin{tabular}{|c|c|c|}\hline
         $\bar{\sigma}$&$\bar{\xi}$&$\mathcal{D}$\\\hline\hline
         0  & $(\infty,\,1,\,1)$ & $\left(0,\,\pm\sqrt{3},\,\mp\sqrt{3}\right)$\\
         $\dfrac{1}{2}$ & $\left(3,\,21+12\sqrt{3},\,21-12\sqrt{3}\right)$ & $\left(\pm\sqrt{2},\,\dfrac{\pm\sqrt{2}}{\sqrt{3}+1},\,\dfrac{\mp\sqrt{2}}{\sqrt{3}-1}\right)$\\
         1 & $(0,\,9,\,9)$ & $\left(\mp2,\,\pm1,\,\pm1\right)$\\
         \hline \hline        
    \end{tabular}
    \label{tab:sigma}
\end{table}

\section{Tetrad fixing in numerical spacetimes}\label{sec:tetrad}

\subsection{Numerical Tetrad}\label{sec:PK}

The simple tetrad we use during evolution is a symmetric null tetrad
constructed from the unit hypersurface normal $\hat{\tau}$ and a set
of three orthonormal unit spatial vectors $\hat{e}_{(1)} =
\hat{e}_{\theta}$, $\hat{e}_{(2)} = \hat{e}_{\phi}$, $\hat{e}_{(3)} =
\hat{e}_r$, suitably orthonormalized via a Gram-Schmidt procedure:
\begin{align}
\label{eqn:OT-null_tetrad}
l^\mu_{\rm num} &\equiv \frac{1}{\sqrt{2}} ( \hat{\tau}^\mu +   \hat{e}_{(3)}^\mu )
\,, 
\cr
n^\mu_{\rm num} &\equiv \frac{1}{\sqrt{2}} ( \hat{\tau}^\mu -
\hat{e}_{(3)}^\mu ) \,, 
\cr
m^\mu_{\rm num} &\equiv \frac{1}{\sqrt{2}} ( \hat{e}_{(1)}^\mu + i \hat{e}_{(2)}^\mu
) \,.
\end{align}

%
Similar tetrads have been commonly used in radiation extraction from
$3+1$ numerical investigations~\cite{Smarr77,Brandt96a,Fiske:2005fx,Sperhake:2005uf,Campanelli:2005ia}, 
and such a tetrad was
used in the earliest investigations of the asymptotic radiative degrees of
freedom of the Weyl tensor~\cite{Szekeres65}. If we have
long-lived 3D numerical evolutions, whose physical domain extends far
from the strong-field region, the $\Psi_4$ extracted should yield a
good measure of the actual outgoing gravitational radiation. We will
refer to Eq.~\eqref{eqn:OT-null_tetrad} hereafter as the \emph{numerical}
tetrad.

Tetrad rotations are classified as Type I, II, and III, and have the
form:
\begin{align}
  \nonumber l^\mu &\to l^\mu \,,\\
  \nonumber n^\mu &\to n^\mu + \bar a m^\mu + a \bar m^\mu + a\bar a l^\mu \,,\\
  \nonumber m^\mu &\to m^\mu+ a l^\mu \,,\\
  \bar m^\mu &\to \bar m^\mu+ \bar a l^\mu \,,
\end{align}  
for Type I, 
\begin{align}
  \nonumber l^\mu &\to l^\mu + \bar b m^\mu + b \bar m^\mu + b\bar b n^\mu \,,\\
  \nonumber n^\mu &\to n^\mu \,,\\
  \nonumber m^\mu &\to m^\mu+ b n^\mu \,,\\
  \bar m^\mu &\to \bar m^\mu+ \bar b n^\mu \,,
\end{align}
for Type II, and
\begin{align}\label{eq:typeIII}
  \nonumber l^\mu &\to A^{-1} l^\mu \,,\\
  \nonumber n^\mu &\to A n^\mu \,,\\
  \nonumber m^\mu &\to e^{iB} m^\mu \,,\\
  \bar m^\mu &\to e^{-iB} \bar m^\mu \,,
\end{align}
for Type III,
where $a$ and $b$ are complex scalars and $A$ and
$B$ are real scalars.

Nerozzi et al.~\cite{Nerozzi:2004wv} give a constructive way to make use of Type I and II tetrad rotations to obtain a {\it transverse frame} in generic spacetimes, i.e., such that $\Psi_1=0$ and $\Psi_3=0$.
Implementation of such strategy has been done in a full numerical context
in Refs.~\cite{Campanelli:2005ia} and~\cite{Zhang:2012ky} for the two main approaches to solve BBH evolutions.

\subsection{Numerical Symmetric Tetrad}\label{sec:ST}

The transverse frame fixes only 4 of 6 tetrad rotation degrees of freedom.
As mentioned they only determine Type I and II rotations, leaving
Type III rotations undetermined. Here, we make use of the symmetric tetrad conditions
in the full nonlinear GR context to fix the Type III rotation parameters.
Here, we use the notations
based on Refs.~\cite{Lousto:2005xu,Campanelli:2005ia}.


Now our symmetric conditions for the spin coefficients read
\begin{align}\label{eq:fixIII}
A^2 &= \dfrac{\rho_{\rm TF}}{\mu_{\rm TF}} \,, \cr
e^{2iB} &= \dfrac{\pi_{\rm TF}}{\tau_{\rm TF}} \,.
\end{align}
With these coefficients we can use the Type III rotation (\ref{eq:typeIII})
to obtain the new symmetric tetrad and compute all the new spin coefficients
and the Weyl scalars. In particular,
\beq
\Psi_4^{S}=\frac{\rho_{\rm TF}\,\tau_{\rm TF}}
{\mu_{\rm TF}\,\pi_{\rm TF}}\,\Psi_4^{\rm TF} \,.
\eeq

In the above way, we completely fixed the tetrad in the full theory and we have a well defined perturbative analogous.
All radiation formulae remain as in the perturbative regime.

For the Kerr metric in Boyer-Lindquist (BL) coordinates, the numerical tetrad --- defined
by Eq.~\eqref{eqn:OT-null_tetrad} with orthonormalized spherical coordinate
directions for the $\hat{e}_{(i)}$ --- takes the form~\cite{Campanelli:2005ia}:
\begin{align}
l^\mu_{\rm num} &= \frac{1}{2} \left\{ \sqrt{\frac{\Omega}{\Delta \Sigma}} ,\,
\sqrt{\frac{\Delta}{\Sigma}} ,\, 0,\, 
\frac{2 a M r}{\sqrt{\Delta \Omega \Sigma}} \right\} \,,
\cr
n^\mu_{\rm num} &= \frac{1}{2} \left\{ \sqrt{\frac{\Omega}{\Delta \Sigma}} ,\, 
- \sqrt{\frac{\Delta}{\Sigma}} ,\, 0,\, 
\frac{2 a M r}{\sqrt{\Delta \Omega \Sigma}} \right\} \,, 
\cr
m^\mu_{\rm num} &= \frac{1}{2} \left\{ 0 ,\, 0,\, \frac{1}{\sqrt{\Sigma}} ,\,  
\frac{i}{\sth} \, \sqrt{\frac{\Sigma}{\Omega}} \right\} \,, 
\label{eqn:num_tetrad2}
\end{align}
where
\begin{align}
\Omega & \equiv \Lambda \, \Sigma + 2 \, M \, a^2 \, r \, \ssth \,,
\cr
\Delta & \equiv \Lambda - 2 \, M \, r \,,
\cr
\Sigma & \equiv r^2 + a^2 \,\ccth 
\cr &= \zeta \bar{\zeta}
\,,
\cr
\Lambda & \equiv r^2 + a^2 \,,
\cr
\zeta & \equiv r + i \, a \, \cth \,.
\end{align}

Such a tetrad will differ strongly from the Kinnersley tetrad; as a
consequence, all Weyl scalars calculated from it will be non-zero. For
the Kerr-BL, these values will be:
\begin{align}
\Psi_0^{\rm num} & = \Psi_4^{\rm num}
\cr
& =  - \frac{M}{2 \Omega \bar{\zeta}^3} [3 ( \Lambda^2 - \Omega )] \,,
\cr
\Psi_1^{\rm num} & = - \Psi_3^{\rm num}
\cr
& =  - \frac{M}{2 \Omega \bar{\zeta}^3}
[3 i \Lambda \sqrt{\Lambda^2 - \Omega}] \,, 
\cr
\Psi_2^{\rm num}
& = - \frac{M}{2 \Omega \bar{\zeta}^3} [- ( 3 \Lambda^2 - \Omega )] \,.
\label{eqn:Weyl_Kerr_num}
\end{align}

\subsection{Kerr Perturbations in the Completely Symmetric Tetrad}\label{sec:Kerr}

We start from the Kinnersley null tetrad:
\bea
l_{\rm K}^\mu &=& \bigg\{ \frac{r^2+a^2}{\Delta},\,
1,\,0,\,\frac{a}{\Delta} \bigg\} \,,
\cr 
n_{\rm K}^\mu &=& \bigg\{ \frac{r^2+a^2}{2\Sigma},\,
-\frac{\Delta}{2\Sigma},\,0,\,\frac{a}{2\Sigma} \bigg\} \,,
\cr
m_{\rm K}^\mu &=& \bigg\{ \frac{ia\sin\theta}{\sqrt{2}\zeta},\,
0,\,\frac{1}{\sqrt{2}\zeta},\,\frac{i}{\sqrt{2}\zeta\sin\theta} \bigg\} \,.
\eea

Using a rotation of Type III
from the Kinnersley tetrad,
we set spin coefficients as $\mu_{\rm S}=\rho_{\rm S}$ and $\pi_{\rm S}=\tau_{\rm S}$.
In these setup, the parameters $A$ and $B$ are obtained as
\begin{align}
A^2 &= \frac{2\Sigma}{\Delta} \,,
\cr 
e^{2iB} &= - \frac{\zeta}{\bar{\zeta}} = - \frac{\zeta^2}{\Sigma} \,,
\end{align}
or
\begin{align}
A &= \sqrt{\frac{2\Sigma}{\Delta}} \,,
\cr
e^{iB} &= i \sqrt{\frac{\zeta}{\bar{\zeta}}} = i \frac{\zeta}{\sqrt{\Sigma}} \,.
\label{eq:boostrotation}
\end{align}
Here, we have picked up only the positive square root of $A^2$ and $\exp(2iB)$.

Then, all spin coefficients
in the completely symmetric tetrad
are explicitly shown as
\begin{align}
\alpha_{\rm S} &= \beta_{\rm S}
\cr &
= \frac{i r\cos\theta+a}{2\sqrt{2 \Sigma}\,\bar{\zeta}\sin\theta} \,,
\cr
\gamma_{\rm S} &= \epsilon_{\rm S}
\cr &
= \frac{Mr -a^2 - iar\cos\theta+iMa\cos\theta}
{2\sqrt{2 \Delta \Sigma}\,\bar{\zeta}} \,,
\cr 
\mu_{\rm S} &= \rho_{\rm S}
\cr &
= - \frac{\sqrt{\Delta}}{\sqrt{2\Sigma}\,\bar{\zeta}} \,,
\cr 
\pi_{\rm S} &= \tau_{\rm S}
= \frac{a\sin\theta}{\sqrt{2\Sigma}\,\bar{\zeta}} \,,
\cr
\kappa_{\rm S} &= \lambda_{\rm S} = \nu_{\rm S} = \sigma_{\rm S} = 0 \,.
\end{align}
It is noted that the case with $A$ in Eq.~\eqref{eq:boostrotation}
and $B=0$ has been discussed in Ref.~\cite{Lousto:2005xu}.
In the curvature scalars, only $\psi_2$ is non-zero and given by
\bea
\Psi_2^{\rm S} = - \frac{M}{\bar{\zeta}^3} \,,
\eea
which is invariant under the Type III rotation.

Next, we discuss the Teukolsky equation~\cite{Teukolsky:1973ha}
for $\Psi_4$ and $\Psi_0$ in the completely symmetric tetrad.
Equation~(2.14) of Ref.~\cite{Teukolsky:1973ha}:
\begin{widetext}
\begin{equation}
\left[(\hat{\Delta} + 3\gamma - \bar{\gamma} + 4\mu + \bar{\mu})
(\hat{D} + 4\epsilon - \rho)
- (\hat{\bar{\delta}} - \bar{\tau} + \bar{\beta} + 3\alpha + 4\pi)
(\hat{\delta} - \tau + 4\beta) - 3\Psi_2^{\rm K}\right]\Psi_4^{\rm K} = 4\pi T_4^{\rm K} \,,
\end{equation}
where the nonindexed differential operators and spin coefficients denote the Kinnersley ones,
changes to
\beq
\left[(\hat{\Delta}_{\rm S} + 3\gamma_{\rm S} - \bar{\gamma}_{\rm S} 
+ 4\mu_{\rm S} + \bar{\mu}_{\rm S})
(\hat{D}_{\rm S} + 4\gamma_{\rm S} - \mu_{\rm S})
- (\hat{\bar{\delta}}_{\rm S} - \bar{\pi}_{\rm S} + \bar{\alpha}_{\rm S}
+ 3\alpha_{\rm S} + 4\pi_{\rm S})
(\hat{\delta}_{\rm S} - \pi_{\rm S} + 4\alpha_{\rm S}) - 3\Psi_2^{\rm S}\right]
\Psi_4^{\rm S} = 4\pi T_4^{\rm S}
\,,
\label{eq:Dp4}
\eeq
and
\beq
\left[(\hat{D} - 3\epsilon + \bar{\epsilon} - 4\rho - \bar{\rho})
(\hat{\Delta} - 4\gamma + \mu)
- (\hat{\delta} + \bar{\pi} - \bar{\alpha} - 3\beta - 4\tau)
(\hat{\bar{\delta}} + \pi - 4\alpha) - 3\Psi_2^{\rm K}\right]\Psi_0^{\rm K} = 4\pi T_0^{\rm K} \,,
\eeq
changes to
\beq
\left[(\hat{D}_{\rm S} - 3\gamma_{\rm S} + \bar{\gamma}_{\rm S} 
- 4\mu_{\rm S} - \bar{\mu}_{\rm S})
(\hat{\Delta}_{\rm S} - 4\gamma_{\rm S} + \mu_{\rm S})
- (\hat{\delta}_{\rm S} + \bar{\pi}_{\rm S} - \bar{\alpha}_{\rm S} 
- 3\alpha_{\rm S} - 4\pi_{\rm S})
(\hat{\bar{\delta}}_{\rm S} + \pi_{\rm S} - 4\alpha_{\rm S}) 
- 3\Psi_2^{\rm S}\right]\Psi_0^{\rm S} = 4\pi T_0^{\rm S}
\,.
\label{eq:Dp0}
\eeq
\end{widetext}
Here, $\hat{\Delta} = n^\alpha \partial_\alpha$,
$\hat{D} = l^\alpha \partial_\alpha$
and $\hat{\delta} = m^\alpha \partial_\alpha$.
$\pi$ in the right hand side of the above equation is
the usual mathematical constant.
In Eqs.~\eqref{eq:Dp4} and \eqref{eq:Dp0}, we have used
the completely symmetric quantities.
We note that only the differential operators and signatures
in front of the spin coefficients are different each other
between Eqs.~\eqref{eq:Dp4} and \eqref{eq:Dp0}.

With the directional derivatives and spin coefficients,
the source terms $T_4$ and $T_0$ change from
\begin{widetext}
\begin{align}
T_4^{\rm K} =& (\hat{\Delta} + 3\gamma - \bar{\gamma} + 4\mu + \bar{\mu})
\left[(\hat{\bar{\delta}} - 2\bar{\tau} + 2\alpha)T_{n\bar{m}}^{\rm K}
- (\hat{\Delta} + 2\gamma - 2\bar{\gamma} + \bar{\mu})T_{\bar{m}\bar{m}}^{\rm K}\right]
\cr &
+ (\hat{\bar{\delta}} - \bar{\tau} + \bar{\beta} + 3\alpha + 4\pi)
\left[(\hat{\Delta} + 2\gamma + 2\bar{\mu})T_{n\bar{m}}^{\rm K}
- (\hat{\bar{\delta}} - \bar{\tau} + 2\bar{\beta} + 2\alpha)T_{nn}^{\rm K}\right] \,,
\end{align}
to
\begin{align}
T_4^{\rm S} =& (\hat{\Delta}_{\rm S} + 3\gamma_{\rm S} - \bar{\gamma}_{\rm S} 
+ 4\mu_{\rm S} + \bar{\mu}_{\rm S})
\left[(\hat{\bar{\delta}}_{\rm S} - 2\bar{\pi}_{\rm S} + 2\alpha_{\rm S})
T_{n\bar{m}}^{\rm S}
- (\hat{\Delta}_{\rm S} + 2\gamma_{\rm S} - 2\bar{\gamma}_{\rm S} 
+ \bar{\mu}_{\rm S})T_{\bar{m}\bar{m}}^{\rm S}\right]
\cr &
+ (\hat{\bar{\delta}}_{\rm S} - \bar{\pi}_{\rm S} + \bar{\alpha}_{\rm S} 
+ 3\alpha_{\rm S} + 4\pi_{\rm S})
\left[(\hat{\Delta}_{\rm S} + 2\gamma_{\rm S} + 2\bar{\mu}_{\rm S})
T_{n\bar{m}}^{\rm S}
- (\hat{\bar{\delta}}_{\rm S} - \bar{\pi}_{\rm S} + 2\bar{\alpha}_{\rm S} 
+ 2\alpha_{\rm S})T_{nn}^{\rm S}\right]
\,,
\label{eq:T4s}
\end{align}
and
\begin{align}
T_0^{\rm K} =& (\hat{D} - 3\epsilon + \bar{\epsilon} - 4\rho - \bar{\rho})
\left[ (\hat{\delta} + 2\bar{\pi} - 2\beta)T_{l m}^{\rm K}
- (\hat{D} - 2\epsilon + 2\bar{\epsilon} - \bar{\rho})T_{mm}^{\rm K} \right]
\cr &
+ (\hat{\delta} + \bar{\pi} - \bar{\alpha} - 3\beta - 4\tau)
\left[(\hat{D} - 2\epsilon - 2\bar{\rho})T_{l m}^{\rm K}
- (\hat{\delta} + \bar{\pi} - 2\bar{\alpha} - 2\beta)T_{ll}^{\rm K}\right] \,,
\end{align}
to
\begin{align}
T_0^{\rm S} =& (\hat{D}_{\rm S} - 3\gamma_{\rm S} + \bar{\gamma}_{\rm S} 
- 4\mu_{\rm S} - \bar{\mu}_{\rm S})
\left[ (\hat{\delta}_{\rm S} + 2\bar{\pi}_{\rm S} - 2\alpha_{\rm S})
T_{l m}^{\rm S}
- (\hat{D}_{\rm S} - 2\gamma_{\rm S} + 2\bar{\gamma}_{\rm S} 
- \bar{\mu}_{\rm S})T_{mm}^{\rm S} \right]
\cr &
+ (\hat{\delta}_{\rm S} + \bar{\pi}_{\rm S} - \bar{\alpha}_{\rm S} 
- 3\alpha_{\rm S} - 4\pi_{\rm S})
\left[(\hat{D}_{\rm S} - 2\gamma_{\rm S} - 2\bar{\mu}_{\rm S})T_{l m}^{\rm S}
- (\hat{\delta}_{\rm S} + \bar{\pi}_{\rm S} - 2\bar{\alpha}_{\rm S} 
- 2\alpha_{\rm S})T_{ll}^{\rm S}\right]
\,,
\label{eq:T0s}
\end{align}
\end{widetext}
respectively. Again, in Eqs.~\eqref{eq:T4s}
and~\eqref{eq:T0s},
we have used the completely symmetric quantities,
and can see the same symmetries shown in Eqs.~\eqref{eq:Dp4}
and~\eqref{eq:Dp0}.

By the way, the separable Teukolsky equations are obtained
for $\Psi_{(s=-2)} = \bar{\zeta}^4 \Psi_4^{\rm K}$
and $\Psi_{(s=2)} = \Psi_0^{\rm K}$
with the help of the ($r,\,\theta$) function, $\bar{\zeta}$.
Here, $s=\pm 2$ is the spin weight of
the perturbed field.
In the Type III rotation, the Weyl scalars
$\Psi_4$ and $\Psi_0$ are transformed as
\begin{align}
\Psi_4 & \to A^2 e^{-2iB} \Psi_4 \,,
\cr
\Psi_0 & \to A^{-2} e^{2iB} \Psi_0 \,.
\end{align}
Therefore, in the completely symmetric tetrad frame,
$\Psi_4$ and $\Psi_0$ are related to the Kinnersley one as
\begin{align}
\Psi_4^{\rm S} &= - \frac{2 \bar{\zeta}^2}{\Delta} \Psi_4^{\rm K} \,,
\cr 
\Psi_0^{\rm S} &= - \frac{\Delta}{2 \bar{\zeta}^2} \Psi_0^{\rm K} \,.
\end{align}
This means that the original Teukolsky equations
with a well-known compact form of the radial derivative,
$\Delta^{-s} \partial_r (\Delta^{s+1} \partial_r)$,
are obtained from $\Psi_4^{\rm S}$ and $\Psi_0^{\rm S}$
by using
\begin{align}
\Psi_{(s=-2)} &= - \frac{\Delta \bar{\zeta}^2}{2} \Psi_4^{\rm S} \,,
\cr
\Psi_{(s=2)} &= - \frac{2 \bar{\zeta}^2}{\Delta} \Psi_0^{\rm S} \,.
\label{eq:SepaSymFun}
\end{align}

About the source term, we can see the consistent expression
in the Kinnersley and completely symmetric tetrads
by using
\begin{align}
T_{nn}^{\rm S} &= \frac{2\Sigma}{\Delta} T_{nn}^{\rm K} \,,
\cr
T_{ll}^{\rm S} &= \frac{\Delta}{2\Sigma} T_{ll}^{\rm K} \,,
\cr
T_{\bar{m}\bar{m}}^{\rm S} &= -\frac{\bar{\zeta}}{\zeta} T_{\bar{m}\bar{m}}^{\rm K} \,,
\cr
T_{mm}^{\rm S} &= -\frac{\zeta}{\bar{\zeta}} T_{mm}^{\rm K} \,,
\cr
T_{n\bar{m}}^{\rm S} &= -i \sqrt{\frac{2}{\Delta}}\bar{\zeta}\,T_{n\bar{m}}^{\rm K} \,,
\cr
T_{lm}^{\rm S} &= i \sqrt{\frac{\Delta}{2}}\frac{1}{\bar{\zeta}}\,T_{lm}^{\rm K} \,.
\end{align}

There remains to explore the fall-off properties of the fields for which we solve the Teukolsky equation
for the completely symmetric tetrad, 
i.e., $\Psi_0^{\rm S}$ and $\Psi_4^{\rm S}$ in order to provide appropriated boundary conditions to solve the corresponding PDEs.
If we may use some factor for $\Psi_0^{\rm S}$ and $\Psi_4^{\rm S}$ as Eq.~\eqref{eq:SepaSymFun},
the differential equations are separable and the original Teukolsky equations.

With regards to an important issue of imposition of boundary conditions
needed for its numerical integration, $\Psi_4$ we can take outgoing boundary conditions
and with $\Psi_0$ ingoing boundary conditions in the Kinnersley tetrad in the frequency/time domain
for large $r$.
For the limit of $r \to \infty$,
we have
\bea
&&\Psi_{(s=-2)} = r^4 \Psi_4^{\rm K} = - \frac{r^4}{2} \Psi_4^{\rm S} \,,
\cr
&&\Psi_{(s=2)} = \Psi_0^{\rm K} = -2 \Psi_0^{\rm S} \,.
\eea
Therefore, we may consider the same boundary condition as those in the Kinnersley tetrad.
Using the Teukolsky equations for $\Psi_4^{\rm S}$ and $\Psi_0^{\rm S}$,
we can see this fact by checking its asymptotic behaviors
with a more detailed analysis given below.
For $\Psi_4^{\rm S}$ and $\Psi_0^{\rm S}$, the asymptotic behaviors will be
\bea
&&\Psi_4^{\rm S} \to \frac{e^{-i \omega (t-r)}}{r} \,~ {\rm and} \,~
\frac{e^{-i \omega (t+r)}}{r^5} \,,
\cr 
&&\Psi_0^{\rm S} \to \frac{e^{-i \omega (t+r)}}{r} \,~ {\rm and} \,~
\frac{e^{-i \omega (t-r)}}{r^5} \,,
\eea
where we have used the $\omega$ mode decomposition
in the frequency domain.

To discuss the asymptotic behaviors for $r^* \to \pm \infty$
(where $dr^*/dr=\Lambda/\Delta$),
we treat the behaviors of $\Psi_{(s=-2)}$ and $\Psi_{(s=2)}$.
First, we use the mode function approach and write the single mode as
\begin{equation}
\Psi_{s} = R_{s}(r)S_{s}(\theta)e^{-i \omega t} e^{i m \phi} \,,
\label{eq:Psi_s}
\end{equation}
where $s=-2$ or $2$, and $R_{s}$ ($S_{s}$) is the radial (spheroidal) wave function
of the radial (angular) Teukolsky equation.
Here, we are omitting the mode indexes ($\ell,\,m,\,\omega$).
Note that 
we can use the same radial $R_s$ and angular $S_s$ functions
given in Eq.~\eqref{eq:Psi_s}
both for the Kinnersley and symmetric tetrads.

The radial mode function, $R_{s}$ is a solution of the radial Teukolsky equation, and
has the asymptotic behaviors,
\bea
&&R_{-2}(r) \to e^{i k r^*} \,~ {\rm and} \,~ \Delta^2 e^{-i k r^*} \,,
\cr
&&R_{2}(r) \to e^{i k r^*} \,~ {\rm and} \,~ \frac{e^{-i k r^*}}{\Delta^{2}} \,,
\eea
for $r^* \to - \infty$, where $k=\omega-ma/(2Mr_+)$
(where $r=r_+$ denotes the outer event horizon),
and
\bea
&&R_{-2}(r) \to r^3 e^{i \omega r^*} \,~ {\rm and} \,~ 
\frac{e^{-i \omega r^*}}{r} \,,
\cr
&&R_{2}(r) \to \frac{e^{i \omega r^*}}{r^{5}} \,~ {\rm and} \,~ \frac{e^{-i \omega r^*}}{r} \,,
\eea
for $r^* \to \infty$ (see, e.g., Ref.~\cite{Teukolsky:1973ha}).
Therefore, in the Kinnersley tetrad we have
\bea
&&\Psi_4^{\rm K} \to e^{i k r^*} \,~ {\rm and} \,~ \Delta^2 e^{-i k r^*} \,,
\cr
&&\Psi_0^{\rm K} \to e^{i k r^*} \,~ {\rm and} \,~ \frac{e^{-i k r^*}}{\Delta^{2}} \,,
\eea
for $r^* \to - \infty$ where there is no change in the asymptotic behaviors
near the horizon, and
\bea
&&\Psi_4^{\rm K} \to \frac{e^{i \omega r^*}}{r} \,~ {\rm and} \,~ 
\frac{e^{-i \omega r^*}}{r^{5}} \,,
\cr
&&\Psi_0^{\rm K} \to \frac{e^{i \omega r^*}}{r^{5}} \,~ {\rm and} \,~ 
\frac{e^{-i \omega r^*}}{r} \,,
\eea
for $r^* \to \infty$.
On the other hand, in the symmetric tetrad we have
\bea
&&\Psi_4^{\rm S} \to \frac{e^{i k r^*}}{\Delta} \,~ {\rm and} \,~
\Delta e^{-i k r^*} \,,
\cr
&&\Psi_0^{\rm S} \to \Delta e^{i k r^*} \,~ {\rm and} \,~ 
\frac{e^{-i k r^*}}{\Delta} \,,
\eea
for $r^* \to - \infty$ where the asymptotic behaviors look symmetric
around the horizon,
and
\bea
&&\Psi_4^{\rm S} \to \frac{e^{i \omega r^*}}{r} \,~ {\rm and} \,~ 
\frac{e^{-i \omega r^*}}{r^{5}} \,,
\cr
&&\Psi_0^{\rm S} \to \frac{e^{i \omega r^*}}{r^{5}} \,~ {\rm and} \,~ 
\frac{e^{-i \omega r^*}}{r} \,,
\eea
for $r^* \to \infty$.

The idea in Ref.~\cite{Lousto:2005xu} is to use combinations of
$\Psi_0\pm\Psi_4$ (in the $m$-mode decomposition) to have separation in even/odd components with the right fall-off (at least in Schwarzschild background, then generalize for Kerr).


\section{Conclusions and discussion}\label{sec:disc}

The question of approximate Petrov types was raised in Ref.~\cite{Campanelli:2008dv} while trying to figure out the nature of the final spacetime product of the merger of two black holes and the possibility of the existence of a transitional Petrov type II-like between the orbital, Petrov type I and final Kerr, Petrov type D spacetimes. In that paper, the direct use of the $\lambda_i$ eigenvalues (\ref{eq:lambdas}) as indicators of the spacetime local algebraic speciality was dependent on the specific numerical tetrad (\ref{eqn:num_tetrad}) 
used. In Ref.~\cite{Owen:2010vw}, a geometrically motivated frame was introduced, but still the lack of a tetrad fixing left the analysis undecided as to the possibility of further exploring the binary black hole merger spacetime.

In this paper, we described a frame in which to study the spacetime, as well as a new method of locally classifying the
approximate Petrov type of the spacetime by the use of the invariant $S$ and index $\mathcal{D}$ in conjunction. These allow us to have a better
idea of the behavior of the spacetime in the strong-field region --- most interestingly near the
black holes and in a region where the spacetime is approximately algebraically special. 
Previously, using only $S$, we could not differentiate between Petrov types II and D in this region. The key observation is that one should use a non-QK frame to transition smoothly from Petrov type I to II to D spacetimes.
This provides us with more insight into the spacetime itself and unique ways to
analyze the strong-field region of the binary black hole system and its merger product. This region, since
such strong dynamical gravitational fields are present, is not particularly well studied, and new methods need
to be developed to accurately analyze what happens near the black holes themselves.

One of the applications of this approach is to full numerical simulations of merging black holes,
with special focus on the latest stage of merger and ringdown. To this end, one can
evaluate first the $S$-invariant, given in Eq.~(\ref{eq:S}), to determine
when the spacetime can be said to be (approximately) algebraic
special and then the $\mathcal{D}$-index, given in Eq.~(\ref{dindex}), to tell when this algebraic
speciality can be characterized by a Petrov type D or II.
Depending on these results being ${\cal D}\sim\pm2$ (or ${\cal D}\sim\pm1$ in a QK frame) one may claim there is a period and region of Petrov type II 
around the merging black holes. If this is the case we can adventure to model the gravitational radiation, 
for instance its power, in terms of a Robinson-Trautman spacetimes \cite{Podolsky:2016sff} and phenomenologically 
model the free functions of the outgoing variable $u$ (see Eq.~(28.26) of Ref.~\cite{Stephani:2003tm}).
One could even add a perturbative solution around these backgrounds \cite{Moreschi:2001qw}.
Another possibility is to use the Chandrasekhar exact algebraic special solutions \cite{Chandrasekhar:1984}, 
as for instance used in Ref.~\cite{Lousto:2002em}, for Kerr perturbations in the case we model a purely 
Petrov type D approach to the final merged black hole.

The algebraic speciality properties described above can be studied by only choosing an appropriated frame. 
Other more specific studies require completely fixing the tetrad. A practical implementation to be applied in
numerically generated spacetimes has been proposed in Ref.~\cite{Nerozzi:2016kky} to match the Kinnersley tetrad. Here we proposed (cf. Eq.~(\ref{eq:fixIII})) a simpler local way to implement such fixing in a symmetric tetrad.


\begin{acknowledgments}
The authors thank J.~Healy and Y.~Zlochower for discussions on the numerical implementation of transverse frames and A.~Nerozzi for clarifications on the reduction of Eq.~(\ref{eq:quartic}) to a quadratic problem.
The authors gratefully acknowledge the National Science Foundation (NSF) for financial support from Grant
No.\ PHY-1912632.
Computational resources were also provided by the NewHorizons, BlueSky Clusters, Green Prairies, and White Lagoon at the Rochester Institute of Technology, which were
supported by NSF grants No.\ PHY-0722703, No.\ DMS-0820923, No.\ AST-1028087, No.\ PHY-1229173, No.\ PHY-1726215,
and No.\ PHY-2018420.
This work used the Extreme Science and Engineering
Discovery Environment (XSEDE) [allocation TG-PHY060027N], which is supported by NSF grant No.\ ACI-1548562. Frontera is an NSF-funded petascale computing system at the Texas Advanced Computing Center (TACC).
H.N. acknowledges support from JSPS KAKENHI Grant No.\ JP21K03582, No.\ 21H01082, and No.\ JP17H06358.
\end{acknowledgments}

\bibliographystyle{apsrev4-2}
\bibliography{references,referencesplus}

\end{document}